\documentclass[%
aps,
prd,
amssymb,
amsmath,
amsfonts,
eqsecnum,
showpacs,
nofootinbib,
floatfix,
twocolumn,
twoside,
a4paper,
superscriptaddress,
longbibliography
]{revtex4-1}

\usepackage{amsmath, latexsym}
\usepackage{amssymb}

\usepackage[T3,T1]{fontenc}
\usepackage{mathrsfs} 
\usepackage{bm} 
\makeatletter
\@ifclassloaded{beamer}
  { 
    \typeout{UsePackages: Detected beamer}
    \usepackage{tgheros}
    
  }
  { 
    \typeout{UsePackages: Did not detect beamer}
    \ifx\asybeamer\undefined 
    \typeout{UsePackages: Detected article}
    \usepackage{tgtermes} 
    \else 
    \typeout{Fonts: Detected asy for beamer}
    \usepackage{tgheros}
    
    \fi
  }
\usepackage{microtype} 
\def\MT@register@subst@font{
  \MT@exp@one@n\MT@in@clist\font@name\MT@font@list
  \ifMT@inlist@\else\xdef\MT@font@list{\MT@font@list\font@name,}\fi}
\makeatother

\usepackage{graphicx} %
\usepackage[x11names,svgnames,rgb]{xcolor} %
\usepackage{xspace}
\usepackage{braket}
\usepackage{accents}
\usepackage{siunitx}
\usepackage{mathtools}
\DeclareSymbolFontAlphabet{\mathrm}{operators}

\definecolor{CiteColor}{rgb}{0.18039, 0.18824, 0.57255}
\definecolor{UrlColor} {rgb}{0.741, 0.173, 0.000}
\definecolor{DarkUrlColor} {rgb}{0.500, 0.110, 0.000}
\definecolor{LinkColor}{rgb}{0.25098, 0.47843, 0.04706}

\makeatletter %
\newcommand{\ShowFont}{%
  \typeout{The main font is \f@encoding \space \f@family \space %
    \f@series \space \f@shape \space at \f@size pt.}%
  \typeout{The math font sizes are \tf@size pt (main), \sf@size pt %
    (script), and \ssf@size pt (scriptscript).}%
  \typeout{The linewidth is \the\linewidth}} %
\makeatother %

\usepackage{xargs}

\usepackage{graphicx}
\usepackage{dcolumn}
\usepackage{bm}
\usepackage{tabularx}
\usepackage{multirow}
\usepackage{capt-of}
\usepackage{color}
\usepackage{url}
\usepackage[utf8]{inputenc}
\usepackage{placeins}
\usepackage{soul}
\usepackage{booktabs}

\usepackage[nolist,nohyperlinks]{acronym}
\usepackage{xspace}
\usepackage[colorlinks]{hyperref}

\usepackage{subfigure}
\usepackage{orcidlink}

\DeclareMathAlphabet{\mathbfsf}{\encodingdefault}{\sfdefault}{bx}{sl}

\newcommand{\be}{\begin{equation}}
\newcommand{\ee}{\end{equation}}
\newcommand{\bea}{\begin{eqnarray}}
\newcommand{\eea}{\end{eqnarray}}

\newcommand{\J}{\mathbf{J}}

\newcommand{\LO}{\mathbf{L_0}}

\definecolor{light-gray}{gray}{0.95}

\definecolor{dodgerblue}{HTML}{1E90FF}
\definecolor{viennared}{HTML}{DA0A14}
\definecolor{ctorange}{HTML}{FF6C0C}
\definecolor{granadagreen}{HTML}{078931}
\definecolor{wales}{HTML}{ff0038}
\definecolor{valenciacfred}{HTML}{ee3524}
\definecolor{barcelonafcgold}{HTML}{edbb00}
\definecolor{jam}{HTML}{A50B5E}
\definecolor{austriawien}{HTML}{441678}

\AtBeginDocument{%
  \hypersetup{
    citecolor=dodgerblue,
    linkcolor=dodgerblue,   
    urlcolor=dodgerblue}}

\newcommand{\UIB}{Departament de F\'isica, Universitat de les Illes Balears, IAC3 -- IEEC, Crta. Valldemossa km 7.5, E-07122 Palma, Spain}

\newcommand{\ICE}
{Institut de Ci\`encies de l'Espai (ICE, CSIC), Campus UAB, Carrer de Can Magrans s/n, 08193 Cerdanyola del Vall\`es, Spain}

\allowdisplaybreaks[1]

\begin{document}

\title[ML]
{Waveform model for the $(\ell=2,m=0)$ spherical harmonic and the displacement memory contribution from precessing binary black holes}


\author{Maria Rossell\'o-Sastre\,\orcidlink{0000-0002-3341-3480}}
\affiliation{\UIB}

\author{Sascha Husa\,\orcidlink{0000-0002-0445-1971}}
\affiliation{\ICE}
\affiliation{\UIB}

\date{\today}

\begin{abstract}
In this paper we construct the first phenomenological waveform model which contains the ``complete'' $\ell=2$ spherical harmonic mode content for gravitational wave signals emitted by the coalescence of binary black holes with spin precession: The model contains the dominant part of the gravitational wave displacement memory, which manifests in the $(\ell=2, m=0)$ spherical harmonic in a co-precessing frame, as well as the oscillatory component of this mode. The model is constructed by twisting up the oscillatory contribution of the mode, as it was previously done for the rest of spherical harmonic modes in {\tt IMRPhenomTPHM} and the Phenom family of waveform models. Regarding the displacement memory contribution present in the aligned spin $(2,0)$ mode, we discuss a procedure to analytically compute the ``precessing memory'' in all the $\ell=2$ modes using the integration derived from the Bondi-Metzner-Sachs balance laws. The final waveform of the $(2,0)$ mode is then obtained by summing together both contributions. We implement this as an extension of the computationally efficient {\tt IMRPhenomTPHM} waveform model, and we test its accuracy by comparing against a set of Numerical Relativity simulations. Finally, we employ the model to perform a Bayesian parameter estimation injection analysis.
\end{abstract}

\maketitle

\section{Introduction}
\label{sec:Introduction}
Gravitational wave (GW) waveform modeling is increasingly focused on developing generic models that incorporate a wide range of physical effects such as spin precession, orbital eccentricity, memory effects, or matter interactions. 
A comprehensive understanding of these effects is essential for refining waveform models employed in GW data analysis and parameter estimation (PE). This necessity becomes particularly significant as the sensitivity of the ground based detector network of LIGO \cite{LIGOScientific:2014pky}, Virgo \cite{VIRGO:2014yos} and KAGRA \cite{KAGRA:2018plz} continues to improve, and next-generation GW observatories, such as the Einstein Telescope (ET) \cite{Abac:2025saz}, the Cosmic Explorer (CE) \cite{CE}, and the Laser Interferometer Space Antenna (LISA) space mission \cite{LISA} are underway. These observatories will enhance the sensitivity to GW signals across diverse frequency bands and enable the detection of novel GW sources. Consequently, it will be necessary to account for previously neglected subdominant physical effects, as their omission could introduce biases in the inferred parameters of the emitting astrophysical sources.

In this study, we present the first phenomenological waveform model for the coalescence of binary black holes (BBH) with misaligned spins, which contains the ``complete'' $\ell=2$ spherical harmonic mode content, in the sense that the model contains the displacement memory, which manifests in the $(\ell=2, m=0)$ spherical harmonic in a co-precessing frame, as well as the oscillatory component of this mode.  This work generalizes our phenomenological waveform model for this mode in the aligned-spin (AS) case \cite{PhysRevD.110.084074}. For the AS case other waveform models have been developed to include the $(2,0)$ spherical harmonic, such as the time domain models {\tt NRHybSur3dq8\_CCE} \cite{PhysRevD.108.064027} and {\tt TEOBResumS-GIOTTO} \cite{Albanesi:2024fts}, and the recent frequency domain model \cite{elhashash2025waveformmodelsgravitationalwavememory} for nonspinning binaries. In addition, the Python package {\tt GWMemory} \cite{PhysRevD.98.064031} provides the code to add the memory correction to various waveform models, including with spin precession, such as {\tt NRSur7dq4} \cite{Varma:2019csw}. 
In this work, we use the Bondi-Metzner-Sachs (BMS) balance laws to compute the displacement memory contribution \cite{Mitman:2020bjf}, and we discuss how the memory and the associated oscillatory component of the signal can be incorporated into waveform models in the context of the twisting-up approximation \cite{PhysRevLett.113.151101, Schmidt:2010it, PhysRevD.86.104063}, which can be used to approximate waveforms of precessing systems by mapping them to AS waveforms. Variations of the  twisting-up approximation have been used widely in modeling precessing waveforms, see e.g. 
\cite{Babak:2016tgq,Ossokine_2020,Pratten:2020ceb,Garcia-Quiros:2020qpx,Estelles:2021gvs}.

There are several different types of memory effects, see for instance Tab.~I from \cite{Mitman:2020bjf}, but from here on, when we mention the memory contribution we refer solely to the displacement memory, since it is the one providing the most significant contribution directly to the strain of the waveform and this is the effect we include in our model.

For nonprecessing binaries, the orbital plane is preserved, and it is therefore natural to take the axis orthogonal to that plane as the reference. In contrast, in systems with precession, the orbital plane itself precesses, so the direction of maximal emission evolves over time. This precession breaks the symmetry present in nonprecessing systems, and consequently the spherical harmonic modes in precessing binaries develop a more complex morphology and become more sensitive to the choice of frame. When performing a rotation of the system, the modes with the same $\ell$ mix among themselves. In systems with no precession, the main memory contributions are contained in the $m=0$ modes (although in nonprecessing, asymmetric systems the memory can be nonzero in some $m\neq0$ modes) \cite{PhysRevD.98.064031, Siddhant:2024nft}. In contrast, in precessing systems, the inertial spherical harmonic modes with $m\neq0$ can also contain significant memory contributions. For this reason, beyond the contribution to the AS $(2,0)$ spherical harmonic, in the precessing case, we must account for the memory contribution in all $\ell=2$ modes.

This article is organized as follows. In Sec. \ref{sec:twisting}, we briefly summarize the twisting-up procedure to obtain gravitational wave signals for precessing systems from their aligned-spin equivalent, and we apply this to the oscillatory component of the $(2,0)$ mode. In Sec. \ref{sec:mem_prec} we discuss the treatment of the memory signal in precessing systems. We recover the memory calculation from the BMS balance laws, review existing waveforms for precessing systems that incorporate the memory effect present in the literature, and introduce an approach to obtain analytical expressions to compute the memory in an inertial frame. In Sec. \ref{sec:implementation} we describe the implementation in the framework of {\tt IMRPhenomTPHM} and we show comparisons with Numerical Relativity (NR). In Sec. \ref{sec:matches} we perform a more systematic study of the accuracy of the model by computing matches against Simulating Extreme Spacetimes (SXS) simulations. In Sec. \ref{sec:pe} we use our waveform model to perform a PE exercise, recovering an injected waveform with Bayesian inference. Sec. \ref{sec:conclusions} summarizes the main conclusions of our work and what can be done in the near future in this line. Finally, in Appendixes \ref{appendix:twistmem} and \ref{appendix:integralinertial} we present the complete analytical expressions developed in this work. Throughout this work, we adopt geometric units ($G=c=1$). The waveforms are rescaled by a factor $R/M$ and the time by $1/M$ in order to match NR units.

\section{Twisting up the oscillatory component}
\label{sec:twisting}

In this Section, we summarize the approximate approach used to map nonprecessing systems to precessing systems, which has commonly been referred to as ``twisting-up'' \cite{PhysRevLett.113.151101, Schmidt:2010it, PhysRevD.86.104063}. This method has been used in previous phenomenological models, and in particular to construct the model we extend in this work, the precessing time domain model {\tt IMRPhenomTPHM} \cite{Estelles:2021gvs}, from the aligned spin model {\tt IMRPhenomTHM} \cite{estelles2020time}. In this procedure, the spherical harmonic modes in the co-precessing frame are taken to be approximately the modes of the corresponding AS system. Then, the modes in an inertial frame can be obtained by performing a 3D time-dependent rotation described by three Euler angles ($\alpha, \beta, \gamma$) from the noninertial co-precessing frame using the Wigner $D$-matrices,
\begin{equation}
\label{rot}
    h_{\ell,m}^{\text{I}} = \sum_{m'=-\ell}^{\ell} \mathcal{D}^{\ell}_{m,m'}(\alpha,\beta,\gamma)h^{\text{cp}}_{\ell,m'}(t).
\end{equation}

\begin{widetext}
\begin{center}
\begin{figure}[htp]
\includegraphics[width=0.74\textwidth]{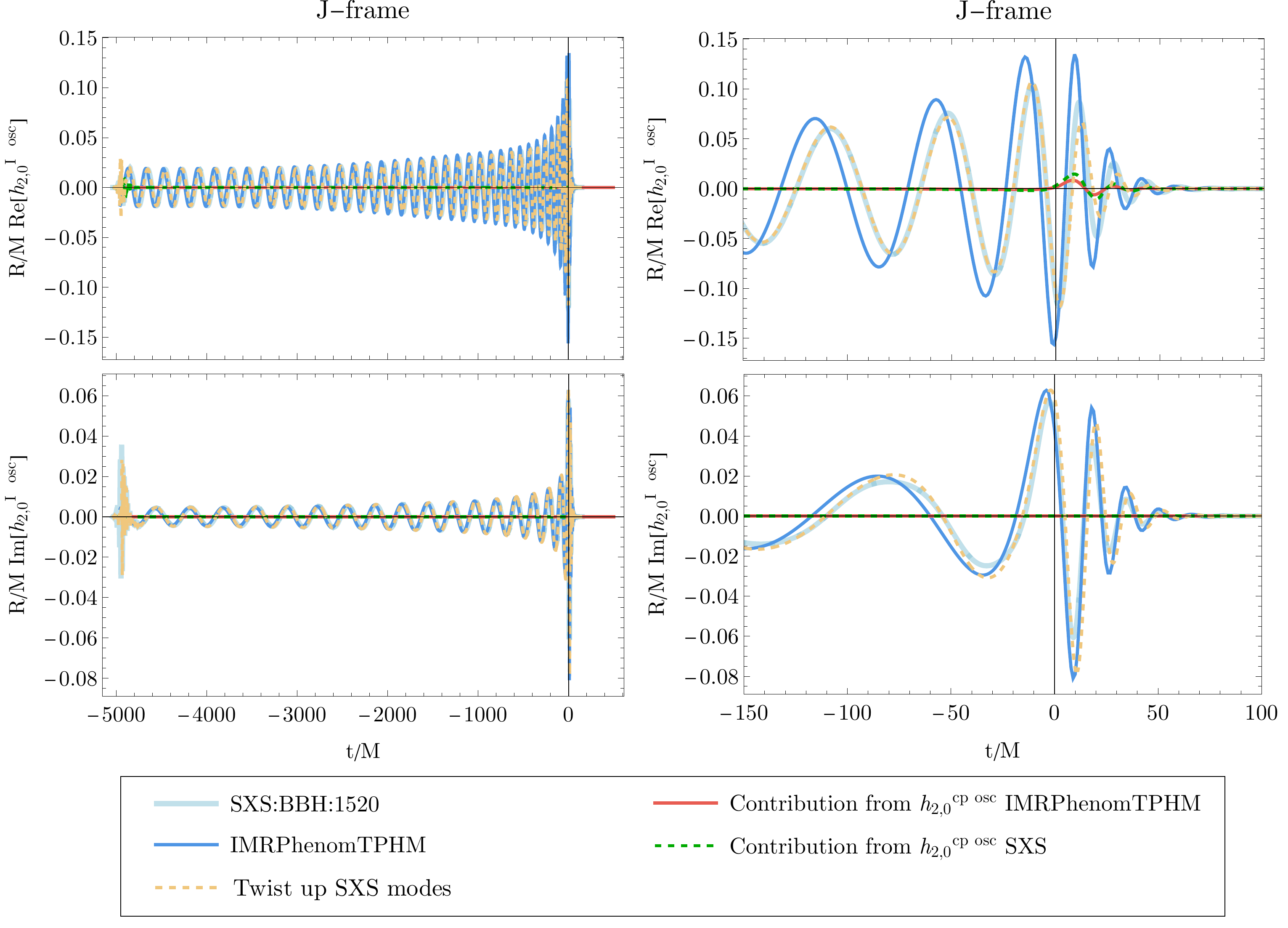}
    \caption{Comparison of our results for the oscillatory contribution of the $(2,0)$ mode in the inertial $\J$-frame with the NR simulation SXS:BBH:1520 with parameters: $q=3.03, \bm{\chi}_1^{\text{ref}}=\{0.540,-0.137,-0.435\}, \bm{\chi}_2^{\text{ref}}=\{0.056,0.258,0.129\}, \chi_{\text{p}}=0.557$ at $Mf_{\text{ref}}=5.15\times10^{-3}$. In light blue, we show the SXS waveform. In solid blue we present the mode computed with {\tt IMRPhenomTPHM}, i.e., using the twisting-up approximation on the co-precessing modes from this model. In dashed yellow, we show the twisting-up of the SXS co-precessing modes. The solid red and dashed green curves show the contribution from the oscillatory component of the $(2,0)$ co-precessing mode for the {\tt IMRPhenomTPHM} model and the SXS simulation, respectively. The top row corresponds to the real part of the mode, and the bottom row to the imaginary part. The left column shows the full evolution of the waveform, and the right column zooms in on the times around the merger.}
    \label{fig:osc_1520}
\end{figure}
\end{center}
\end{widetext}

Here and in the rest of this article, the superscripts ``I'' and ``cp'' stand for inertial and co-precessing, respectively. We follow the same conventions as those used in the {\tt IMPRhenomTPHM} model, see \cite{Estelles:2021gvs} for details. The Euler angles in an inertial frame are defined as
\begin{subequations}
\begin{equation}
    \alpha = \arctan(\hat{L}_y/\hat{L}_x),
\end{equation}
\begin{equation}
\label{beta_angle}
    \cos\beta = \hat{\bm{z}}\cdot\hat{\bm{L}}=\hat{L}_z,
\end{equation}
\begin{equation}
    \dot{\gamma} = -\dot{\alpha}\cos\beta,
\end{equation}
\end{subequations}
where $\bm{L}$ is the orbital angular momentum. The angles $\alpha$ and $\beta$ approximately track the direction of the Newtonian angular momentum, and the third angle $\gamma$ is fixed by enforcing the minimal rotation condition \cite{PhysRevD.84.124011}. Here and throughout the rest of this work, the over-dot indicates the time derivative. In Eq.~(\ref{rot}) we observe how the modes with the same $\ell$ mix when performing the rotation. 

The ``twisted'' $(2,0)$ mode as a function of the Euler angles and all the $\ell=2$ modes can be written as
\begin{equation}
\label{twist_osc}
\begin{split}
    h_{2,0}^{\text{I}}&=\frac{1}{4} \left\{h_{2,0}^{\text{cp}}\left[1 + 3 \cos (2 \beta )\right]\right.\\
    &+2 \sqrt{6} \left[\sin ^2(\beta) \left(\text{Re}\left[h_{2,2}^{\text{cp}}\right] \cos (2 \gamma ) + \text{Im}\left[h_{2,2}^{\text{cp}}\right] \sin (2 \gamma )\right)\right.\\
    &\left.\left.-i \sin (2 \beta )\left(\text{Re}\left[h_{2,1}^{\text{cp}}\right] \sin (\gamma) - \text{Im}\left[h_{2,1}^{\text{cp}}\right] \cos (\gamma)\right)\right]\right\}.
\end{split}
\end{equation}

We use this expression to obtain the oscillatory contribution of the inertial $(2,0)$ spherical harmonic mode. We replace the $h_{\ell,m}^{\text{cp}}$ modes by the AS modes from {\tt IMRPhenomTHM}. In the case of $h_{2,0}^{\text{cp}}$, we use only the AS oscillatory contribution, excluding the memory. The reason why we do not simply twist up the full AS $(2,0)$ mode including the memory contribution is that the twisting-up procedure and the memory integration do not commute, as will be explained in Sec. \ref{sec:twistmem}. Consequently, it becomes necessary to follow a separate procedure to twist up the memory contribution to obtain the full precessing $(2,0)$ spherical harmonic mode.

In Fig.~\ref{fig:osc_1520} we provide a plot to visually assess the accuracy of the twisting-up procedure for the $(2,0)$ spherical harmonic mode using the SXS:BBH:1520 simulation from the general SXS catalog \cite{SXS:catalog}, which includes only the oscillatory contribution of the mode, but not the memory. We start the waveform at the initial frequency of the SXS simulation, $f_{\text{ini}}$, and we use as a reference frequency for the {\tt IMRPhenomTPHM} model the reference frequency of the SXS simulation, $f_{\text{ref}}$. We set the reference value of the spin components at this reference frequency.
By comparing the yellow dashed curves (corresponding to the twisting-up of the SXS modes) with the light blue curves (NR data), we can check the accuracy of the twisting-up procedure; while comparing light blue and solid blue curves, we assess the accuracy of the twisting-up implementation using the {\tt IMRPhenomTPHM} model in comparison to NR. The solid red curves show the contribution of the oscillatory component of the $(2,0)$ co-precessing mode for the model, which is obtained by twisting up the AS mode.  We employ the numerical evolution of the precessing spin equations (using the flag {\tt PhenomXPrecVersion=300} in the model). The dashed green curves show the same, but for the SXS simulation, for which we use the Euler angles obtained when rotating the modes from the co-precessing frame to the $\J$-frame, which is the inertial frame aligned with the total angular momentum of the binary system $(\text{\textbf{J}}=\text{\textbf{L}}+\text{\textbf{S}}_1+\text{\textbf{S}}_2)$. As expected, the imaginary part of this contribution vanishes, as the mode $h_{2,0}^{\text{cp}}$ is real. Near the peak of the real part of the mode, we observe a dephasing of the model waveform with respect to the NR simulation. This discrepancy may arise from a loss of accuracy of the Euler angles of the model at this stage of the binary evolution.

\section{Memory contribution for precessing systems}
\label{sec:mem_prec}
Once the oscillatory signal of the $(2,0)$ spherical harmonic mode has been treated using the twisting-up approximation applied to the AS waveform, the remaining contribution — namely the displacement memory — must be computed. It is important to emphasize that, in precessing systems, the memory is distributed among all the modes, but we focus on the $\ell=2$ modes for simplicity, as they contain the dominant contribution of the displacement memory. Each mode must be computed independently, to this end, we employ the expression derived from the BMS balance laws, as detailed in Sec. \ref{sec:framework}.

\subsection{Availability of precessing waveforms containing the memory effect}
The SXS catalog of NR simulations contains two public simulations of precessing binary black holes that explicitly incorporate the memory effect. These waveforms are obtained using the Cauchy Characteristic Extraction (CCE) method implemented in the SpECTRE code and are available in the Ext-CCE waveform catalog \cite{SXS:ExtCCE, SXSCCEarticle}.

The first simulation (SXS:BBH\_ExtCCE:0008), corresponds to an equal-mass system, where the objects have the following initial spins $\bm{\chi}_1^{\text{ini}}=(0.487, 0.125, -0.327), \bm{\chi}_2^{\text{ini}}=(-0.190, 0.051, -0.227)$, and evolves over 20.47 orbits, starting at $Mf_{\text{ini}}=4.38\times10^{-3}$. The second simulation (SXS:BBH\_ExtCCE:0013) corresponds to a system with mass ratio $q=4$ and the same initial spins as the previous simulation, but in this case, it evolves over 17.43 orbits, beginning at $Mf_{\text{ini}}=5.64\times10^{-3}$. Comparing both simulations, the memory is more prominent for equal-mass systems, as in the AS case, since this is the case where the GW emission is maximized.

In addition to these two simulations, with built-in memory contribution, it is also possible to add the memory correction to the waveforms in the general SXS Catalog \cite{SXS:catalog, sxs_mem} via the {\tt sxs.waveforms.memory.add\_memory} functionality. We utilize this approach to be able to extend our comparison with NR data beyond the two Ext-CCE cases. Since these baseline waveforms lack memory by default, they also provide a means to validate the analytical prescriptions used to reconstruct the memory contribution from waveforms that do not contain it.

Moreover, memory effects can be incorporated into precessing waveforms using the {\tt GWMemory} package \cite{PhysRevD.98.064031}, which has been used to add the memory to the NRSurrogate precessing waveform model, {\tt NRSur7dq4} \cite{Varma:2019csw}.

\subsection{Framework for computing the displacement memory}
\label{sec:framework}
As discussed in \cite{PhysRevD.110.084074}, the displacement memory contribution can be obtained from the energy flux term in the supermomentum balance law in the set of BMS balance laws, as shown in \cite{Mitman:2020pbt,Mitman:2020bjf}, which is given by
\begin{equation}
\label{je}
    J_{\varepsilon}=\frac{1}{2}\bar{\eth}^2\mathfrak{D}^{-1}\left[\frac{1}{4}\int_{u_1}^u\left|\dot{h}\right|^2 du\right]+\alpha(\theta,\phi),
\end{equation}
where $\alpha(\theta,\phi)$ is an undetermined function on the sphere that accounts for the difference between the strain computed from the balance laws, $J$, and the true strain $h$ due to possible ambiguities in the choice of Bondi frame.
The spin-weight operators act on spin-weighted spherical harmonics as
\begin{equation}
    \bar{\eth}^sY_{\ell,m}=-\sqrt{(\ell+s)(\ell-s+1)}\;^{s-1}Y_{\ell,m},
\end{equation}
and, on $s=0$ spherical harmonics,
\begin{equation}
    \mathfrak{D}Y_{\ell,m}=\frac{1}{8}(\ell+2)(\ell+1)\ell(\ell-1)Y_{\ell,m},
\end{equation}
as defined in \cite{Mitman:2020pbt}.

To simplify the calculation of the memory, we decompose the gravitational wave strain into a basis of $s=-2$ spin-weighted spherical harmonics as
\begin{equation}
\begin{split}
    h(t,r,\theta_{\text{JN}},\phi)&=h_+-ih_{\times}\\
    &=\frac{1}{r}\sum_{\ell=2}^{\infty}\sum_{m=-\ell}^{m=\ell}h_{\ell,m}(t)\;^{-2}Y_{\ell,m}(\theta_{\text{JN}},\phi),
\end{split}
\end{equation}
where
\begin{equation}
    ^sY_{\ell,m}(\theta_{\text{JN}},\phi)=(-1)^s\sqrt{\frac{2\ell+1}{4\pi}}\;^{-s}d_{\ell,m}(\theta_{\text{JN}})e^{im\phi},
\end{equation}
with 
\begin{widetext}
\begin{equation}
    ^sd_{\ell,m}(\theta_{\text{JN}})=\sum_{k=\max(0,m-s)}^{\min(\ell+m,\ell-s)}(-1)^k\frac{\sqrt{(\ell+m)!(\ell-m)!(\ell+s)!(\ell-s)!}}{(\ell+m-k)!(\ell-s-k)!k!(k+s-m)!}\cos^{2\ell+m-s-2k}\left(\frac{\theta_{\text{JN}}}{2}\right)\sin^{2k+s-m}\left(\frac{\theta_{\text{JN}}}{2}\right),
\end{equation}
\end{widetext}
and the angular coordinates $(\theta_{\text{JN}},\phi)$ are related to Cartesian coordinates in the standard way.

In the AS case, the dominant displacement memory contribution is confined to the $(2,0)$ mode. However, for precessing systems analyzed in an inertial frame, contributions from all $\ell=2$ modes must be considered, as they can also exhibit non-negligible memory signatures.

\subsection{Twist up the co-precessing modes and compute the memory}
\label{sec:twistmem}
In this Section, we aim to develop analytical expressions for computing the memory contribution of the inertial $\ell=2$ modes from the co-precessing modes. We first apply the twisting-up procedure to the $\ell=2$ modes, assuming equatorial symmetry ($h^*_{\ell,m}=(-1)^{\ell}\;h_{\ell,-m}$). For illustrative purposes, we consider the simplification that the nondominant $\ell=2$ modes do not contribute to the other modes in the twisting up approximation. The full, correct expressions including all the $\ell=2$ modes are presented in Eqs.~(\ref{full_expr}). We compute the expression for the memory in Eq.~(\ref{je}) and project the result onto each of the $\ell=2$ modes. Applying the corresponding operators $\left(\frac{1}{8}\bar{\eth}^2\mathfrak{D}^{-1}\right)$, we get the following expressions for the memory contribution to each mode in the inertial frame
\begin{subequations}
\label{twistmem}
\begin{equation}
\label{twistmem20}
        h_{2,0}^{\text{I\;mem}}=\frac{1}{28} \sqrt{\frac{5}{6 \pi }} \int_{u_1}^u(1+3 \cos (2 \beta )) \left|\dot{h}_{2,2}^{\text{cp}}\right|^2 du,
\end{equation}
\begin{equation}
        h_{2,2}^{\text{I\;mem}}=\frac{1}{28} \sqrt{\frac{5}{\pi }}\int_{u_1}^u e^{-2 i \alpha } \sin ^2(\beta)\left|\dot{h}_{2,2}^{\text{cp}}\right|^2 du,
\end{equation}
\begin{equation}
        h_{2,-2}^{\text{I\;mem}}=\frac{1}{28} \sqrt{\frac{5}{\pi }}\int_{u_1}^u e^{2 i \alpha } \sin ^2(\beta)\left|\dot{h}_{2,2}^{\text{cp}}\right|^2 du,
\end{equation}
\begin{equation}
        h_{2,1}^{\text{I\;mem}}=-\frac{1}{14} \sqrt{\frac{5}{\pi }}\int_{u_1}^u e^{-i \alpha } \sin(\beta) \cos(\beta)\left|\dot{h}_{2,2}^{\text{cp}}\right|^2 du,
\end{equation}
\begin{equation}
        h_{2,-1}^{\text{I\;mem}}=\frac{1}{14} \sqrt{\frac{5}{\pi }}\int_{u_1}^u e^{i \alpha } \sin(\beta) \cos(\beta)\left|\dot{h}_{2,2}^{\text{cp}}\right|^2 du.
\end{equation}
\end{subequations}
Comparing Eq.~(\ref{twistmem20}) with the expression for the AS equivalent of the $(2,0)$ spherical harmonic mode, given by
\begin{equation}
\label{h20AS}
    h_{2,0}^{\text{AS\;mem}}=\frac{1}{7} \sqrt{\frac{5}{6 \pi }} \int_{u_1}^u \left|\dot{h}_{2,2}^{\text{AS}}\right|^2 du,
\end{equation}
we see that since $|\cos(2\beta)|\leq 1$, precession always reduces the amplitude of the memory relative to the AS $(2,0)$ mode. This behavior is expected, as mode mixing during twisting-up redistributes the memory across all $\ell=2$ modes.

In this derivation, we have neglected time derivatives of the Euler angles. In this case, where we compute the memory contribution neglecting the nondominant $\ell=2$ modes in the twisting-up approximation, we check that the slow variation of the Euler angles across the time evolution does not lead to a loss of accuracy. We find that the terms containing the time derivatives give a negligible contribution compared to the main term that contains the time derivative of the $(2,2)$ mode. The full expression for the $(2,0)$ mode memory contribution, including these derivatives, can be found in Eq.~(\ref{h20memeuler}). In Fig.~\ref{fig:comp_mem_terms}, we compare the memory waveform with and without the inclusion of the Euler angle derivatives for three different SXS simulations. The final memory amplitude differs by $0.02\%$, $0.12\%$, and $0.44\%$, respectively. For all tested cases, the difference between the two approaches is negligible. Therefore, we choose to neglect these terms in the following for simplicity.

As previously mentioned, in these results we have assumed that the $|m|\neq2$ co-precessing modes do not contribute to the other modes in the twisting-up. If we generalize it to all the $\ell=2$ modes, we obtain the expressions presented in Eqs.~(\ref{full_expr}), in Appendix \ref{appendix:twistmem}, which are the ones we use for the comparisons in the following Section.

Alternatively, we could take a different approach to compute the memory in an inertial frame. Instead of applying the twisting-up first, we could first compute the memory in the co-precessing frame and then twist up the modes to rotate them to the inertial frame. However, as we now demonstrate, this procedure is not equivalent to the previous approach, since the twisting-up and the integration operations do not commute. In this case, we first compute the memory in the $(2,0)$ mode in the co-precessing frame and then we compute the ``twisted'' memory using Eq.~(\ref{rot}). Since the only contribution comes from $h_{2,0}^{\text{cp\;mem}}$, the resulting expressions are:

\begin{subequations}
\label{memtwist}
\begin{equation}
    h_{2,0}^{\text{I\;mem}}=\frac{1}{4} (1+3 \cos (2 \beta ))h_{2,0}^{\text{cp\;mem}},
\end{equation}
\begin{equation}
    h_{2,2}^{\text{I\;mem}}=\frac{1}{2} \sqrt{\frac{3}{2}} e^{-2 i \alpha }  \sin ^2(\beta) h_{2,0}^{\text{cp\;mem}},
\end{equation}
\begin{equation}
    h_{2,-2}^{\text{I\;mem}}=\frac{1}{2} \sqrt{\frac{3}{2}} e^{2 i \alpha } \sin ^2(\beta) h_{2,0}^{\text{cp\;mem}},
\end{equation}
\begin{equation}
    h_{2,1}^{\text{I\;mem}}=-\sqrt{\frac{3}{2}} e^{-i \alpha } \sin(\beta)\cos(\beta) h_{2,0}^{\text{cp\;mem}},
\end{equation}
\begin{equation}
    h_{2,-1}^{\text{I\;mem}}=\sqrt{\frac{3}{2}} e^{i \alpha } \sin(\beta)\cos(\beta) h_{2,0}^{\text{cp\;mem}}.
\end{equation}
\end{subequations}

\begin{figure}[htp]
\includegraphics[width=1\columnwidth]{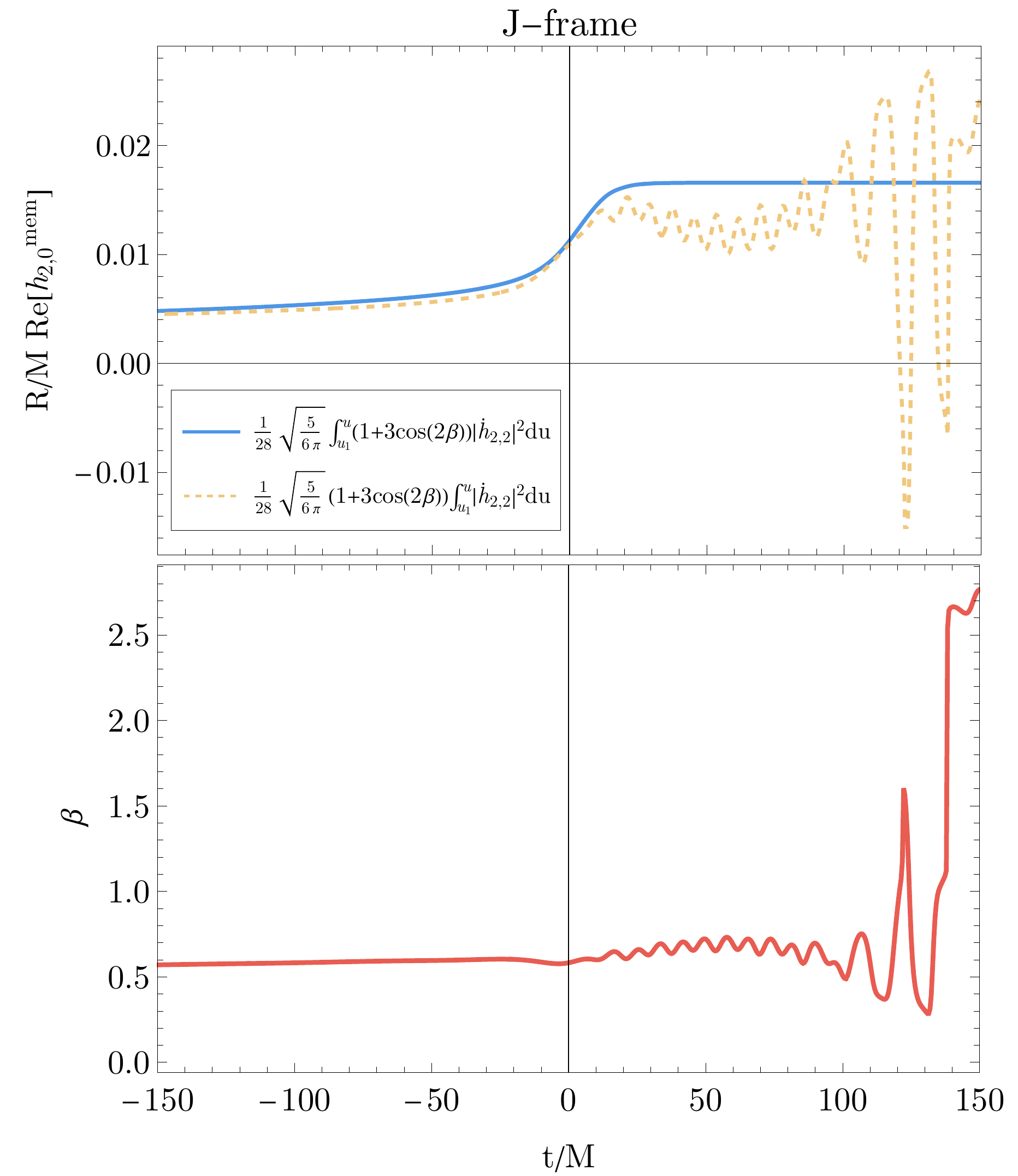}
    \caption{Top panel: comparison between the two methods to compute the memory contribution in the inertial frame. The solid blue curve is the correct result, which consists of twisting up the modes and then performing the time integral, whereas the yellow dashed curve is the incorrect result, where we first compute the memory and then twist up the result. Bottom panel: time evolution of the Euler angle $\beta$, corresponding to the rotation from the co-precessing frame to the inertial $\J$-frame, computed using Eq.~(\ref{beta_angle}). This corresponds to a system with parameters: $q=3.03, \bm{\chi}_1^{\text{ref}}=\{0.540,-0.137,-0.435\}, \bm{\chi}_2^{\text{ref}}=\{0.056,0.258,0.129\}, \chi_{\text{p}}=0.557$ at $Mf_{\text{ref}}=5.15\times10^{-3}$.}
    \label{fig:compmethods_beta}
\end{figure}

As we assume that we can approximate the spherical harmonic modes in the co-precessing frame by the ones of the corresponding AS system, here we approximate $h_{2,0}^{\text{cp\; mem}}$ by the strain of the AS $(2,0)$ mode \cite{PhysRevD.110.084074} ($h_{2,0}^{\text{cp\;mem}}\approx h_{2,0}^{\text{AS\;mem}}$), which is given by Eq.~(\ref{h20AS}). If we replace this expression into Eqs.~(\ref{memtwist}), we do not recover Eqs.~(\ref{twistmem}), because, in the latter, the time integration only applies to the absolute value of the derivative of the $(2,2)$ mode, while the term involving the Euler angles is not affected. This contrasts with the previous expressions, where the time integration encompasses the whole expression, including the terms with the Euler angles. Since the angles evolve slowly during the inspiral regime, both methods give very similar waveforms in this part of the time evolution. However, during the ringdown, the angles can experience more abrupt variations, leading to unphysical oscillations at late times in the evolution, which deviate from the correct result obtained using the first method. Therefore, the inconsistency arises not from the accuracy of the twisting-up approximation but from the incorrect assumption in the second calculation on the commutativity of the memory integration and the twisting-up operation. Hence, the memory integration must be applied to the modes in the inertial frame.

To illustrate this, Fig.~\ref{fig:compmethods_beta} shows the result of the real part of the memory contribution in the $(2,0)$ spherical harmonic computed with both methods in the top panel, and the time evolution of the angle $\beta$, computed within the {\tt IMRPhenomTPHM} model, between the co-precessing frame and the $\J$-frame in the bottom panel, both for times near the merger. We select this mode because it is the one that presents the most significant contribution, therefore, the difference between the two results is more noticeable. We clearly observe that the yellow dashed curve, which corresponds to the second (wrong) method, presents growing oscillations in the ringdown, caused by the term involving the angle $\beta$ and the fact that it is incorrectly placed outside the time integration. This also explains why we cannot simply apply the twisting-up procedure to the full AS $(2,0)$ spherical harmonic mode to get its precessing equivalent. Instead, the oscillatory and memory parts of the waveform must be treated separately.

\subsection{Integration of the inertial modes}
\label{sec:integralinertial}
The first derivation presented in the previous Section is useful for understanding, in a simplified way, how the operations of twisting-up and memory integration should be handled. However, for simplicity, we made two key assumptions: equatorial symmetry of the modes, and slow variation of the Euler angles throughout the evolution, such that their time derivatives can be neglected. While this previous method provides reasonably accurate waveforms as a first approximation, a more general approach, free from these simplifying assumptions, can be derived and will be presented in this Section.

\begin{widetext}
\begin{center}
\begin{figure}[htp]
\includegraphics[width=0.73\textwidth]{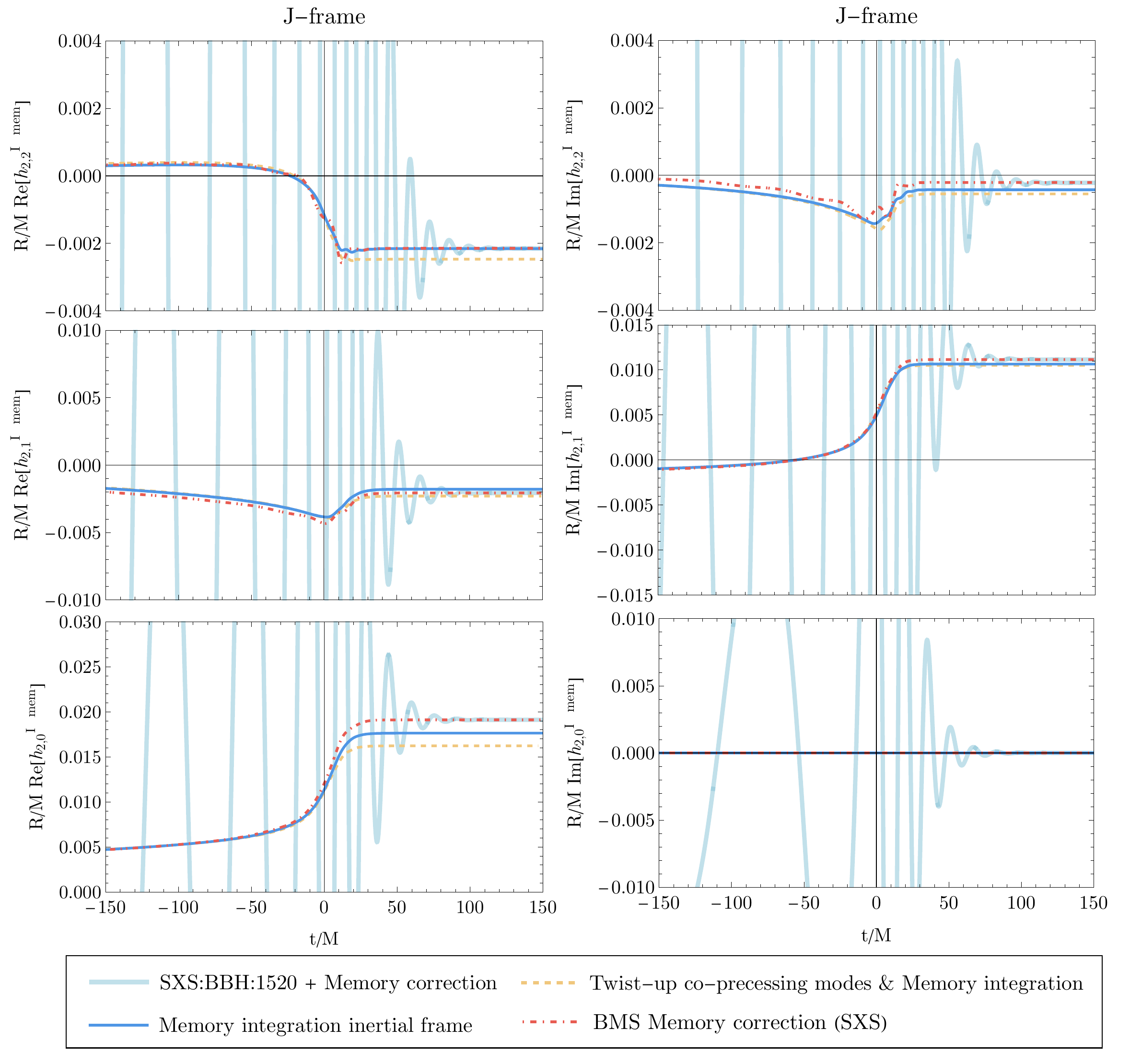}
    \caption{Comparison of our results for the memory contribution the inertial $\J$-frame with the NR simulation SXS:BBH:1520 with parameters: $q=3.03, \bm{\chi}_1^{\text{ref}}=\{0.540,-0.137,-0.435\}, \bm{\chi}_2^{\text{ref}}=\{0.056,0.258,0.129\}, \chi_{\text{p}}=0.557$ at $Mf_{\text{ref}}=5.15\times10^{-3}$. In light blue, we show the complete waveform with the memory correction added from the BMS balance laws (red dash-dotted line). In dashed yellow, we present the memory contribution computed by twisting up the co-precessing modes and then calculating the memory integral. In solid blue it is shown the memory contribution computed using the memory integration of the modes in the inertial $\J$-frame. The left column corresponds to the real part of the $\ell=2$ modes and the right column to the imaginary part.}
    \label{fig:compNR1520}
\end{figure}
\end{center}
\end{widetext} 

In the following, we introduce an alternative formulation in which the memory contribution is computed directly in the inertial frame. This method does not neglect the time derivatives of the Euler angles, does not rely on the twisting-up approximation, and does not impose equatorial symmetry on the modes. Specifically, we allow for the most general case in which $h^*_{\ell,m}\neq (-1)^{\ell\;}h_{\ell,-m}$, treating the $m<0$ modes as independent quantities in the calculations. 

Although the inertial-frame modes provided by the {\tt IMRPhenomTPHM} waveform model are constructed using the twisting-up approximation under the assumption of equatorial symmetry, the expressions derived here can be applied to arbitrary inertial-frame modes, including those obtained from models that do not adopt these approximations. However, in this work, we apply the method to the {\tt IMRPhenomTPHM} modes; thus, the results become equivalent to those of the previous method, except that in the present method, the time derivatives of the Euler angles are included.

To derive the expressions, we first construct the inertial strain using all the $\ell=2$ modes, and then we compute the memory contribution in each of the inertial modes using Eq.~(\ref{je}), and project the result onto the basis of spin-weighted spherical harmonics again. The complete expressions that are obtained can be found in Eqs.~(\ref{expr_intertial}) in Appendix \ref{appendix:integralinertial}. The treatment of the constant of integration $\alpha(\theta,\phi)$ in this calculation is explained in Sec. \ref{sec:implementation}.

Fig.~\ref{fig:compNR1520} presents a comparison between this method and the first one introduced in the previous Section. This comparison serves to validate the consistency of the analytical expressions developed in both procedures. To this end, we employ an NR simulation (SXS:BBH:1520), excluding any built-in memory correction, to ensure a clean comparison. As before, we start the waveform at $f_{\text{ini}}$, corresponding to the initial frequency of the SXS simulation and we use as a reference frequency for the {\tt IMRPhenomTPHM} model the reference frequency of the SXS simulation, $f_{\text{ref}}$. The spin components are specified at this reference frequency. The dashed yellow curve represents the first method in the previous Section, which consists of twisting up the co-precessing modes to have them in the inertial $\J$-frame and then computing the memory integration. In solid blue, we show the memory integration of the inertial modes in the $\J$-frame, as explained in this Section. We then add the memory correction to the simulation using the implementation of the BMS balance laws in the code, which is shown in dash-dotted red lines. The full waveform, incorporating both contributions, is depicted in light blue. We show the comparisons at times near the merger (as the memory amplitude grows suddenly at the merger, whereas during the inspiral regime, all waveforms remain consistent) for all the $\ell=2, m\geq0$ modes. The imaginary part of the $(2,0)$ mode does not contain any contribution from the memory; it is just composed of the oscillatory signal. Both procedures give consistent results among themselves and also with NR. The first method (dashed yellow curves) differs a bit more from NR, especially in the case of the real part of the $m=0$ mode, but the overall behavior is reproduced by both of our analytical methods.

As previously stated, the difference between the two methods mainly arises from the treatment of the time derivatives of the Euler angles. In particular, we have assumed these derivatives to be negligible in the first method, when deriving Eqs.~(\ref{twistmem}). However, discrepancies between the resulting waveforms are observed during the merger-ringdown phase in Fig.~\ref{fig:compNR1520}, where such terms may contribute more significantly. The observed inconsistencies might originate from the assumptions made in the first derivation. Consistency between the methods has been verified when only the dominant $(2,\pm2)$ modes are included in the calculation (as shown in Fig.~\ref{fig:comp_mem_terms}), but this may no longer hold when incorporating the remaining $\ell=2$ modes. Due to the increasing complexity of the resulting analytical expressions, these have not been explicitly derived. Given that the inertial-frame modes can be directly integrated, we adopt this approach, as it provides greater accuracy by avoiding the assumptions made in the other method. Furthermore, the {\tt IMRPhenomTPHM} waveform model directly outputs the inertial-frame modes, allowing for a straightforward and precise computation of the memory contribution. Therefore, the method that has been implemented is the one corresponding to Eqs.~(\ref{expr_intertial}).

\section{Model implementation}
\label{sec:implementation}
We have implemented the precessing $(2,0)$ mode and the memory contribution in all the $\ell=2$ modes within the {\tt IMRPhenomTPHM} waveform model \cite{Estelles:2021gvs} in its Python implementation, {\tt phenomxpy} \cite{phenomxpy}. The memory contribution in all the modes is computed using the time integration of the modes in the inertial frame, following the equations presented in Appendix \ref{appendix:integralinertial}, and using the {\tt cumulative\_trapezoid} method as implemented in {\tt SciPy} \cite{scipy}. On the other hand, the oscillatory part of the $(2,0)$ mode is obtained by twisting up the AS model of this mode, i.e., Eq.~(\ref{twist_osc}). We use {\tt PhenomXPrecVersion=300}, which employs the numerical evolution of the precessing spin equations in {\tt IMRPhenomTPHM}. While this version is slower than approximations such as the next-to-next-to-leading order (NNLO) effective single-spin approximation \cite{Bohe_2013, Marsat_2014} or the double-spin multiscale analysis (MSA) \cite{PhysRevD.95.104004}, it provides the highest available accuracy and remains computationally efficient for practical use. We have introduced configurable options in the code to enable or disable components of the $(2,0)$ mode in the co-precessing frame. When this mode is enabled, the user can select among the following configurations:

\begin{itemize}
    \item Include the full mode (oscillatory and memory contributions) together with the memory contribution in the rest of the $\ell=2$ modes.
    \item Include only the full $(2,0)$ mode but without the memory in the other modes.
    \item Include only the memory contribution in the $(2,0)$ mode.
    \item Include only the oscillatory contribution of this mode.
\end{itemize}

\begin{widetext}
\begin{center}
\begin{figure}[htp]
\includegraphics[width=1\textwidth]{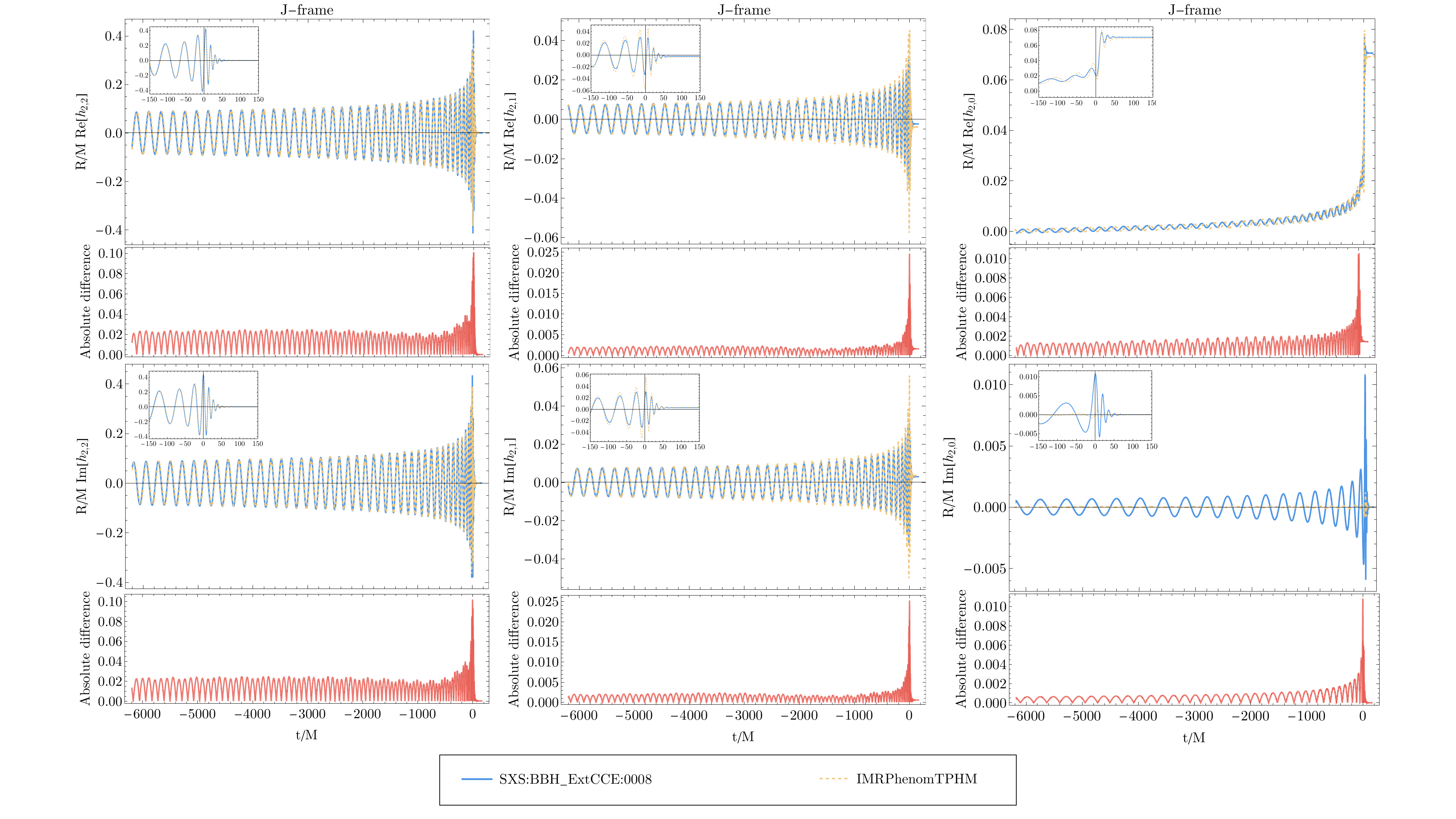}
    \caption{Comparison of the full  $\ell=2, m\geq0$ spherical harmonic modes from the NR simulation SXS:BBH\_ExtCCE:0008 (solid blue) with the {\tt IMRPhenomTPHM} model (dashed yellow) in the inertial $\J$-frame. The first row shows the real part of the modes, while the third row shows the imaginary part. The insets show a zoom in on the times of the merger-ringdown stage of the evolution. The second and fourth rows, in red, stand for the waveform difference between NR and the model, in the modes on top of them. The mismatch between NR and our model for this waveform, computed as in Eq. (\ref{mismatch}), is $6.1\times10^{-2}$.}
    \label{fig:comp_CCE_8}
\end{figure}

\begin{figure}[htp]
\includegraphics[width=1\textwidth]{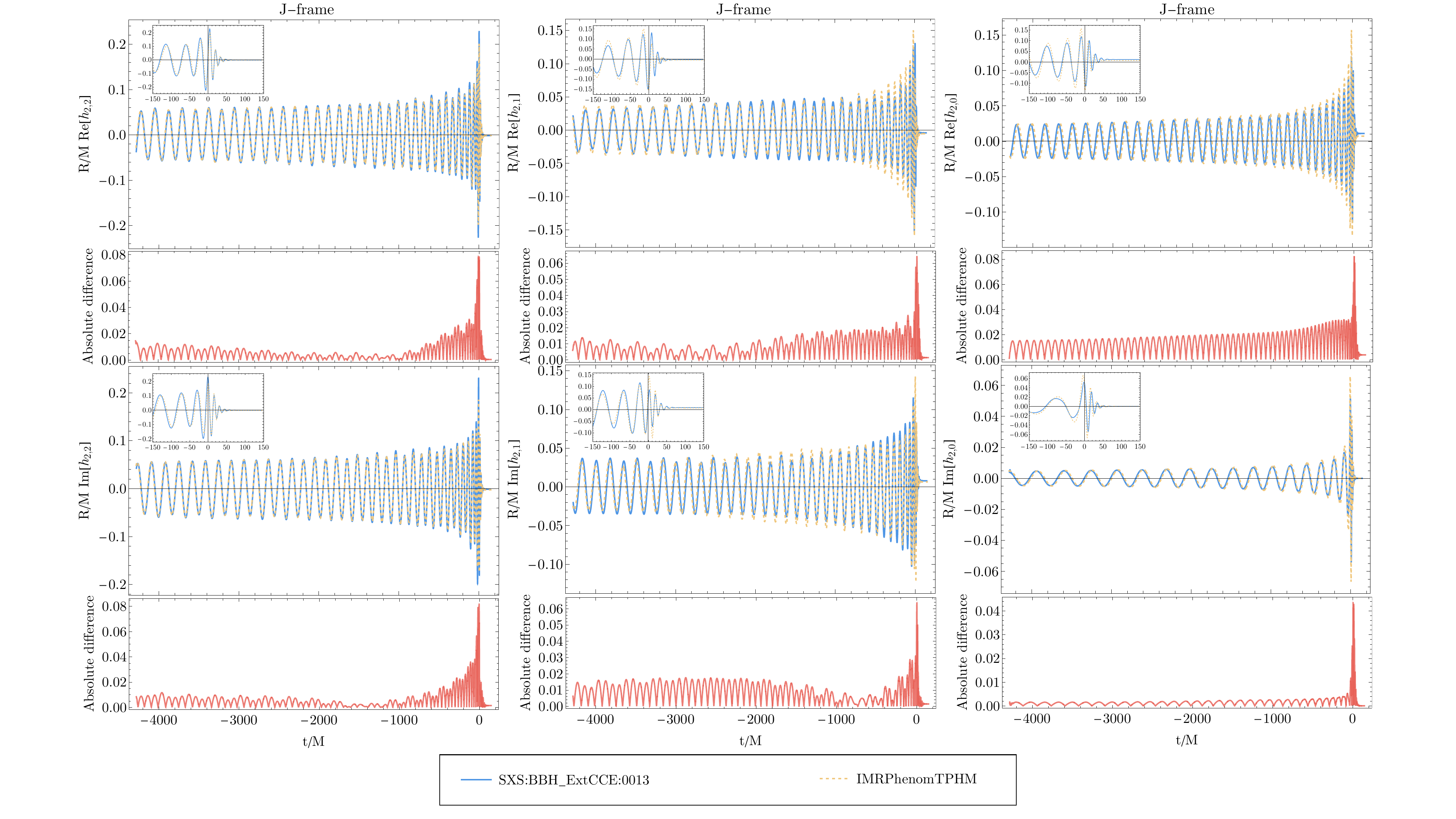}
    \caption{Same as Fig.~\ref{fig:comp_CCE_8}, but for SXS:BBH\_ExtCCE:0013.The mismatch between NR and our model for this waveform, computed as in Eq. (\ref{mismatch}), is $1.2\times10^{-3}$.}
    \label{fig:comp_CCE_13}
\end{figure}
\end{center}
\end{widetext}

Figs. \ref{fig:comp_CCE_8} and \ref{fig:comp_CCE_13} present comparisons between the full $\ell=2, m\geq0$ spherical harmonic modes obtained with our model and the two NR simulations of precessing systems from the SXS:BBH\_ExtCCE catalog. Our results demonstrate that we successfully recover the final memory offset in all the cases, as well as the full $(2,0)$ mode after summing the oscillatory and memory contributions. There are some small discrepancies between the waveforms due to the fact that we employ the twisting-up approximation. Although in Fig.~\ref{fig:comp_CCE_13} we see that the full $(2,0)$ is well recovered, in the case of the equal-mass simulation in Fig.~\ref{fig:comp_CCE_8}, the imaginary part of this mode is recovered with a significantly smaller amplitude. A systematic study quantifying the accuracy of the model is presented in Sec.~\ref{sec:matches}.

As seen in Eq. (\ref{je}), the memory obtained by integrating the oscillatory modes contains an arbitrary additive constant $\alpha(\theta,\phi)$, corresponding to the choice of Bondi frame at the start of the integration. In the results shown we did not perform an explicit BMS-frame alignment: the integration constant in the model was set to zero by construction when integrating the modes. To quantify the impact of the arbitrary integration constant in the memory integral when comparing to NR data, we performed a \textit{post-hoc} check: we computed the time averages of the NR memory and of the model memory over an early interval after the junk radiation region and extracted the required offset. For SXS:BBH\_ExtCCE:0008, we find offsets of 0.34\% for the $(2,2)$ mode, 0.23\%, for the $(2,1)$ mode, and 0.25\%, for the $(2,0)$ mode relative to the peak of the memory amplitude; and for SXS:BBH\_ExtCCE:0013, the corresponding values are of 0.30\%, 0.10\%, and 0.34\% for each of the modes. These offsets are small (below 0.50\% of the peak memory amplitude), so for simplicity the integration constant is kept at zero, and the plots in Figs. \ref{fig:comp_CCE_8} and \ref{fig:comp_CCE_13} confirm that the waveforms are consistent under this choice. The small magnitude of the offsets indicates that this does not significantly affect the accuracy of the waveform model. While incorporating this additional parameter in the match optimization would likely yield improved agreement with NR, we maintain the simpler approach for clarity.

To assess the computational performance, we compute the mean evaluation time needed to generate the waveform polarizations in the $\LO$-frame including the $\ell=2$ modes for the different options implemented in the model across a range of total masses between 10 and 100 $M_{\odot}$, with $f_{\text{ini}}=f_{\text{ref}}=10$ Hz and a sampling frequency of 4096 Hz. To obtain the waveform in the $\LO$-frame, the rotation from the $\J$-frame is performed as implemented in the {\tt IMRPhenomTPHM} model.

\begin{figure}[htp]
\includegraphics[width=1\columnwidth]{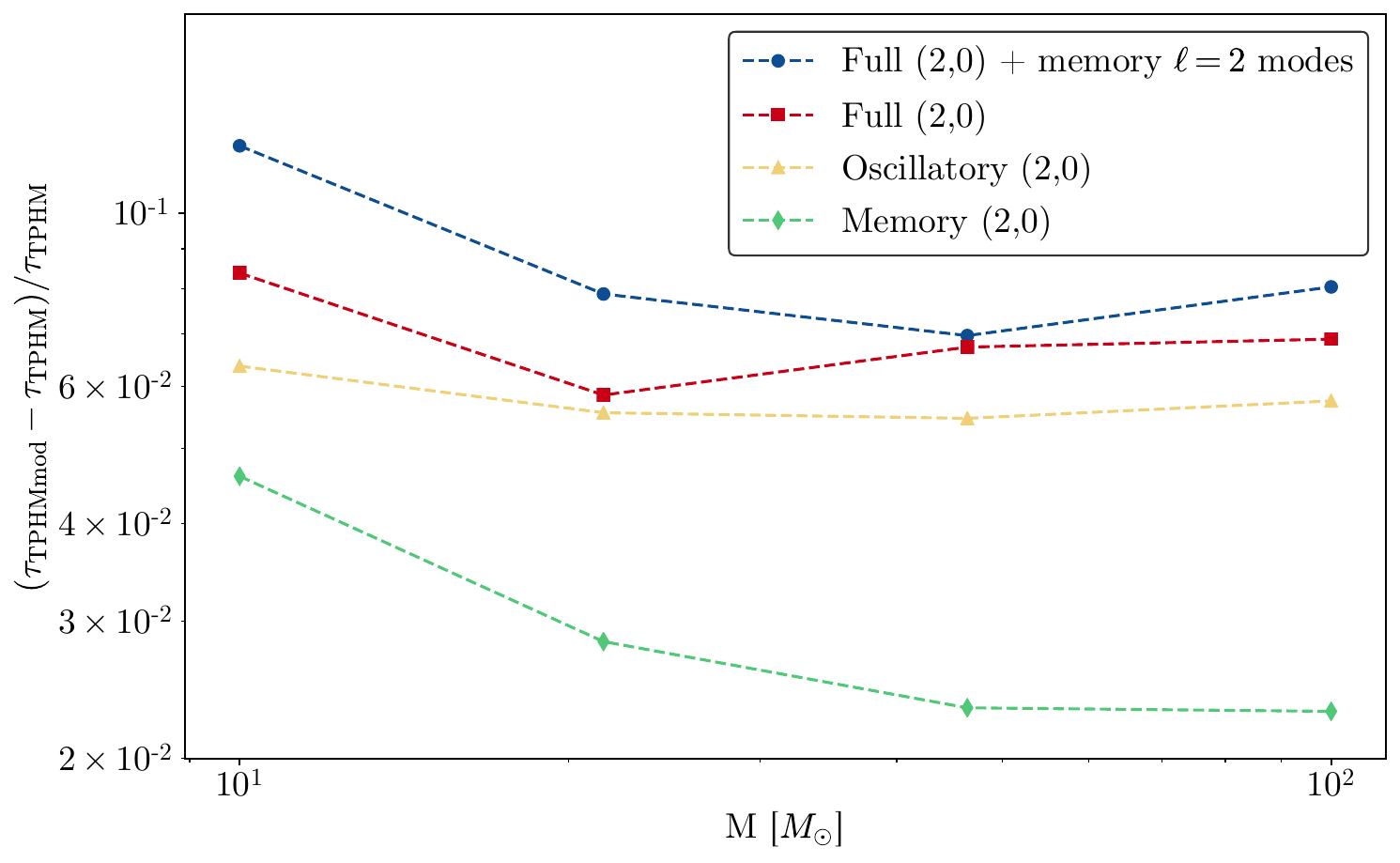}
    \caption{Relative difference in the mean evaluation time for different model versions relative to the baseline model as a function of the total mass. The baseline to which we compare corresponds to the generation of the {\tt IMRPhenomTPHM} model in time domain with all the higher modes in the co-precessing frame included:  $(2,\pm2), (2,\pm1), (3,\pm3), (4,\pm4)$ and $(5,\pm5)$. $\tau_{\text{TPHM}}$ stands for the mean evaluation time of the baseline, while $\tau_{\text{TPHMmod}}$ stands for the baseline model modified by adding each of the features listed in the legend of the plot. This corresponds to an equal-mass system with spins: $\bm{\chi}_1^{\text{ini}}=(0.487, 0.125, -0.327), \bm{\chi}_2^{\text{ini}}=(-0.190, 0.051, -0.227)$.}
    \label{fig:ev_times}
\end{figure}

In Fig.~\ref{fig:ev_times}, we present the relative change in the mean waveform evaluation time when modifying the baseline {\tt IMRPhenomTPHM} model. The baseline includes the co-precessing modes $(2,\pm2)$, $(2,\pm1)$, $(3,\pm3)$, $(4,\pm4)$, and $(5,\pm5)$. We compare this with four modified configurations: the addition of the complete $(2,0)$ mode along with the memory contribution across all $\ell=2$ modes (blue dots), the complete $(2,0)$ mode only (red squares), the oscillatory component of the $(2,0)$ mode only (yellow triangles), and the memory component of the $(2,0)$ mode only (green diamonds). We denote the average evaluation time of the baseline model by $\tau_{\text{TPHM}}$, and that of each modified configuration by $\tau_{\text{TPHMmod}}$. The quantity plotted, $(\tau_{\text{TPHMmod}} - \tau_{\text{TPHM}})/\tau_{\text{TPHM}}$, thus quantifies the relative computational cost of each modification with respect to the baseline. 

The results indicate that the additional computational cost introduced by including the $(2,0)$ mode and memory effects is not significant and only weakly dependent on waveform duration. The inclusion of only the memory component of the $(2,0)$ mode results in the smallest overhead, while the simultaneous inclusion of the full $(2,0)$ mode and memory across all $\ell=2$ modes incurs the highest cost. Nonetheless, the overhead remains limited in all cases, as the most demanding computation is the evaluation of the baseline model, demonstrating that the incorporation of these additional features does not significantly compromise the computational efficiency of the model.

\section{Matches against Numerical Relativity}
\label{sec:matches}
To quantitatively evaluate the accuracy of the model, we compare the waveforms against the corresponding NR waveform from the SXS Catalog using the mismatch calculation \cite{Bruegmann:2007bri}, which is defined as
\begin{equation}
\label{mismatch}
    1-\mathcal{M}=1-\underset{\Delta t, \Delta \varphi}{\max}\frac{\langle h|g\rangle}{\sqrt{\langle h|h\rangle \langle g|g\rangle}},
\end{equation}
where the usual definition of the inner product between two waveforms, weighted by the noise estimate from the power spectral density (PSD) of the detector, is given by
\begin{equation}
    \langle h|g\rangle\equiv 4 \text{Re}\left[\int_{f_{\text{min}}}^{f_{\text{max}}}\frac{\tilde{h}(f)\tilde{g}^*(f)}{S_n(f)}df\right].
\end{equation}
The overlap is maximized over the relative time and phase shifts, meaning that it is computed after aligning the two waveforms in both time and phase. A mismatch value close to zero therefore indicates a high level of agreement between the two waveforms.

\begin{figure}[htp]
\includegraphics[width=0.9\columnwidth]{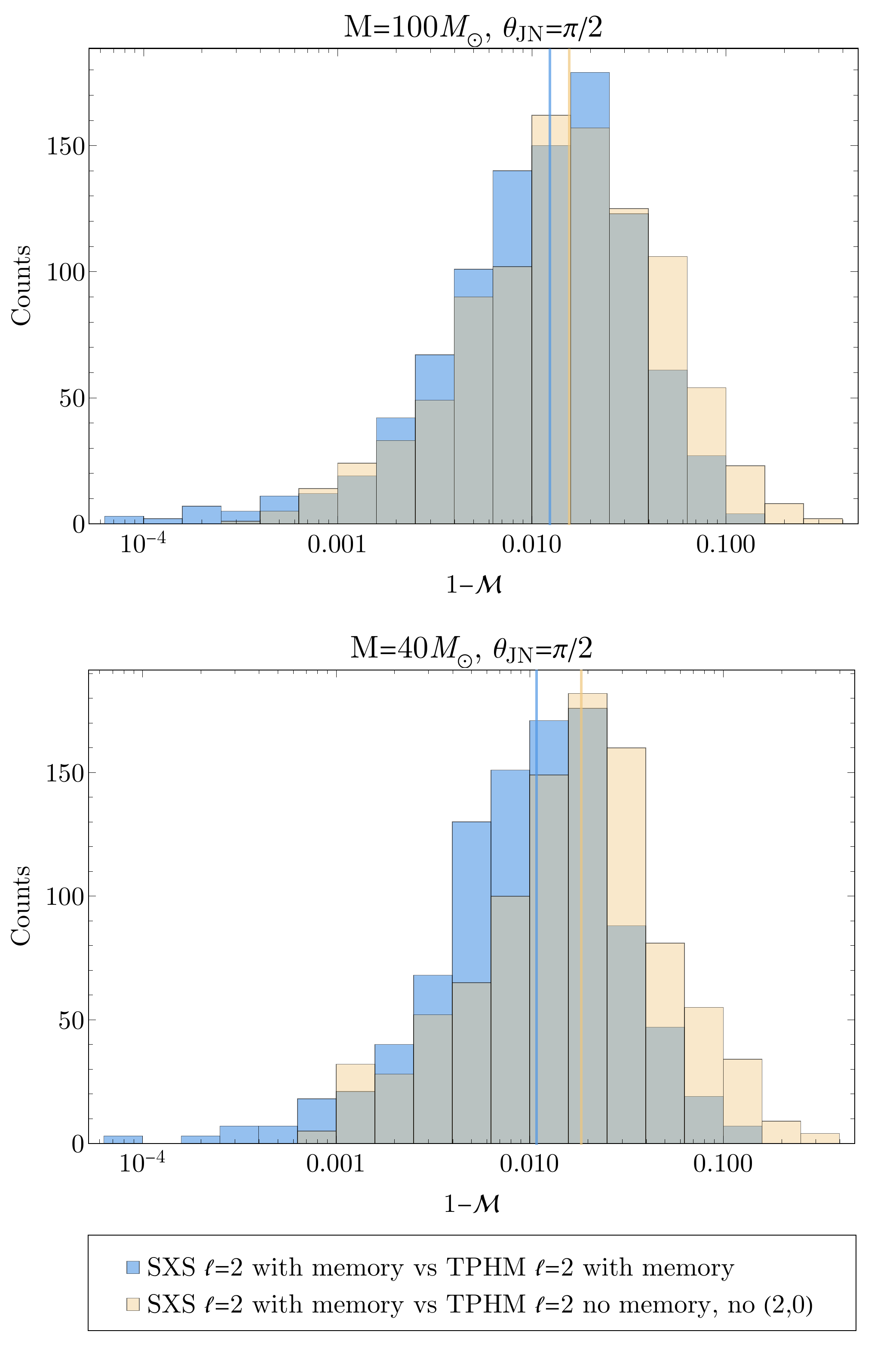}
    \caption{Distributions of mismatches between NR and the {\tt IMRPhenomTPHM} model for a set of 1076 SXS simulations from the general catalog, adding the memory correction. The top panel stands for a total mass of 100$M_{\odot}$ and the bottom panel for a 40$M_{\odot}$, both at an inclination of $\pi/2$ using LIGO A$^\#$ sensitivity from \cite{Sensitivitycurves}. The blue histograms represent the comparison between both waveforms containing all $\ell=2$ modes plus the memory contribution, while the yellow histograms correspond to the SXS waveform with the same mode content as before, but the model waveform neglecting the $(2,0)$ mode in the co-precessing frame and the memory contribution in all the modes. The vertical lines represent the median value of the distribution of the same color.}
    \label{fig:match}
\end{figure}

We select a set of 1076 quasi-circular, precessing simulations from the general SXS Catalog, to which we subsequently add the memory contribution using the {\tt sxs.waveforms.memory.add\_memory} option implemented in the SXS package. For these calculations, we use the PSD of the LIGO A$^\#$ detector, from \cite{Sensitivitycurves}, and we consider the modes in the inertial $\J$-frame. In Fig.~\ref{fig:match}, we present the results for $M=100M_{\odot}$ (top panel) and $M=40M_{\odot}$ (bottom panel) in an edge-on configuration ($\theta=\pi/2$), since this is the inclination that enhances the $(2,0)$ spherical harmonic contribution. The blue histograms in the plots represent the mismatch distribution between SXS and {\tt IMRPhenomTPHM} when both waveforms include all the $\ell=2$ modes in the co-precessing frame and the memory contribution in each of them. The yellow histograms show the mismatch distribution where the SXS waveforms still include all the $\ell=2$ modes and memory, but the model waveform excludes the $(2,0)$ mode contribution in the co-precessing frame and also neglects the memory component in all the modes. This analysis allows us to quantify the improvement in the accuracy of the model when these features are incorporated into the waveform. The incorporation of these new features in the waveform model shifts the mismatch distributions toward lower values, reflecting improved agreement with NR results. Although these are subdominant effects added to the waveform, this result indicates an overall improvement in accuracy, yielding reasonable mismatch values between NR and the model. Additionally, we find a greater improvement for lower total mass, as for $M=40M_{\odot}$ the median mismatch improves from $1.8\times10^{-2}$ to $1.1\times10^{-2}$, whereas for $M=100M_{\odot}$, the median values are $1.6\times10^{-2}$ and $1.2\times10^{-2}$, respectively. This behavior is expected, as lower mass systems produce longer signals that enter the sensitive band of the detectors at lower frequencies. In this frequency regime, the memory contribution is more significant due to its characteristic $1/f$ dependence in the Fourier domain. Consequently, neglecting this effect as well as the contribution of the $(2,0)$ mode leads to higher mismatches for lower masses.

\section{Bayesian Parameter Estimation}
\label{sec:pe}
In the first place, we provide a brief overview of PE principles. According to Bayes' theorem, the probability distribution of the model parameters given the observed data is expressed as
\begin{equation}
    \label{bayes}
    p(\Theta|d)=\frac{\mathcal{L}(d|\Theta)\pi(\Theta)}{\mathcal{Z}},
\end{equation}
where $\mathcal{L}(d|\Theta)$ represents the likelihood of obtaining the data $d$ for a given set of parameters $\Theta$, $\pi(\Theta)$ is the prior distribution that reflects any existing knowledge about the parameters before analyzing the data, and $\mathcal{Z}$ is a normalization constant defined as
\begin{equation}
    \label{normfactor}
    \mathcal{Z}=\int d\Theta \mathcal{L}(d|\Theta)\pi(\Theta).
\end{equation}
The prior distribution $\pi(\Theta)$ encodes the information we have about the parameters before observing the data, while the likelihood function $\mathcal{L}(d|\Theta)$ quantifies the agreement between the observed data and the model predictions. For Gaussian noise, the likelihood function can be written as
\begin{equation}
\label{likelihood}
    \mathcal{L}(d|\Theta)=\exp\left(-\frac{1}{2}\langle h_{\text{inj}}-h(\Theta)| h_{\text{inj}}-h(\Theta) \rangle\right),
\end{equation}
where $\langle \cdot | \cdot \rangle$ denotes the noise-weighted inner product and $h_{\text{inj}}$ is the injected signal.

As illustrated in Fig.~\ref{fig:ev_times}, the efficiency of the model remains competitive after incorporating the developments introduced in this work. Thus, we can employ the model to perform PE studies without incurring excessive computational costs.

To assess the impact of the $(2,0)$ spherical harmonic mode and the displacement memory contribution on the estimation of source parameters in precessing systems, we perform an injection-recovery study in zero-noise in the LIGO A$^{\#}$ sensitivity \cite{Sensitivitycurves} accounting for both Hanford (H1) and Livingston (L1) detectors. We investigate potential parameter biases that arise when neglecting these features by injecting a signal that includes the $(2,0)$ mode and the memory contribution in all the $\ell=2$ modes and then attempting to recover the parameters using both this same version of the model and the baseline {\tt IMRPhenomTPHM} that omits them.

For this purpose, we use the {\tt phenomxpy} implementation, along with the Bilby code \cite{bilby_paper} with {\tt bilby\_pipe} \cite{bilby_pipe_paper} and the nested sampling algorithm {\tt dynesty} \cite{dynesty}. To compute the Fourier transform of the nonperiodic signals, accounting for the step induced by the displacement memory, we employ the {\tt gw-foutstep} method \cite{PhysRevD.110.124026}. 

In Tab.~\ref{tab:priors}, we outline the priors used for the sampled parameters. The mass parameters are given in the detector frame. We denote the masses of the individual components by $m_i$, such that the total mass of the system is $M=m_1+m_2$. The chirp mass is defined as $\mathcal{M}=\frac{(m_1m_2)^{3/5}}{M^{1/5}}$ and the mass ratio, $q=m_2/m_1\leq1$. The dimensionless spin magnitudes are given by $a_i=\frac{|\text{\textbf{S}}_i|}{m_i^2}$. The tilt angles describe the orientation of each spin vector relative to the orbital momentum and are given by $\theta_i=\cos^{-1}\left(\frac{\text{\textbf{S}}_i·\text{\textbf{L}}}{|\text{\textbf{S}}_i||\text{\textbf{L}}|}\right)$. The in-plane spin angle  describes the relative azimuthal angle between the two spin vectors in the plane perpendicular to the orbital angular momentum, $\phi_{12}=\cos^{-1}\left(\frac{(\text{\textbf{S}}_1\times\text{\textbf{L}})·(\text{\textbf{S}}_2\times\text{\textbf{L}})}{|\text{\textbf{S}}_1\times\text{\textbf{L}}||\text{\textbf{S}}_2\times\text{\textbf{L}}|}\right)$. The orbital spin angle is the azimuthal angle between the total angular momentum $(\text{\textbf{J}}=\text{\textbf{L}}+\text{\textbf{S}}_1+\text{\textbf{S}}_2)$ and the orbital angular momentum $\bm{L}$, $\phi_{\text{JL}}=\cos^{-1}\left(\frac{(\text{\textbf{J}}\times\text{\textbf{L}})·(\text{\textbf{S}}_{\text{eff}}\times\text{\textbf{L}})}{|\text{\textbf{J}}\times\text{\textbf{L}}|·|\text{\textbf{S}}_{\text{eff}}\times\text{\textbf{L}}|}\right)$, with $\text{\textbf{S}}_{\text{eff}}=\frac{m_1^2\text{\textbf{S}}_{1}+m_2^2\text{\textbf{S}}_{2}}{m_1^2+m_2^2}$. We present the posteriors for the effective spin parameter \cite{Ajith:2009bn, PhysRevD.78.044021,PhysRevD.82.064016}, defined as
\begin{equation}
    \chi_{\text{eff}}=\frac{m_1a_1\cos\theta_1+m_2a_2\cos\theta_2}{M},
\end{equation}
which describes the dominant nonprecessing spin component; and the effective spin precession parameter \cite{PhysRevD.91.024043}
\begin{equation}
    \chi_{\text{p}}=\frac{\max(A_1S_{1\perp},A_2S_{2\perp})}{A_1m_1^2},
\end{equation}
where $A_1=2+3/2q$, $A_2=2+3q/2$ and $S_{i\perp}=|\text{\textbf{L}}\times(\text{\textbf{S}}_i\times\text{\textbf{L}})|$. This variable quantifies the in-plane spin components and is bounded within $0\leq\chi_{\text{p}}\leq1$. The larger $\chi_{\text{p}}$ is, the stronger the system's precession. These two quantities, $\chi_{\text{eff}}$ and $\chi_{\text{p}}$, are sufficient to describe the spin effects of the binary system.
The angle between the total angular momentum vector and the line of sight between the binary and the detector is referred to by $\theta_{\text{JN}}$, and the luminosity distance is represented as $d_L$. The sky location of the binary is specified using equatorial coordinates: right ascension, denoted by $\alpha$, and declination, denoted by $\delta$. The polarization angle is denoted by $\psi$, and the coalescence time and reference phase are given by $t_c$ and $\phi$, respectively. In the sampling procedure, in order to reduce the computational cost, we keep fixed the parameters: $\alpha, \delta, \psi$ and $\phi$. Regarding the mass components, the individual component masses are uniformly distributed. Instead of directly applying a uniform prior to $q$ and $\mathcal{M}$, the samples for $m_{1,2}$ are drawn from a uniform distribution within the range stated in the constraint prior. For the luminosity distance, we use the Uniform Source Frame prior, which is uniform in comoving volume and source frame time. 

\begin{table}[h!]
\centering
\begin{tabular}{ccccc}
\toprule
\multicolumn{1}{l}{\textbf{Variable}} & \multicolumn{1}{l}{\textbf{Unit}} & \textbf{Prior}        & \textbf{Minimum} & \textbf{Maximum} \\ \hline \hline
$m_{1,2}$                             & $M_\odot$                         & Constraint               & 10                & 100              \\
$q$                                   & -                                 & Uniform in Components & 0.125            & 1                \\
$\mathcal{M}$                         & $M_\odot$                         & Uniform in Components & 15               & 100              \\
$a_{1,2}$                             & -                                 & Uniform               & 0                & 0.99             \\
$\theta_1$                            & rad                               & Sin                   & 0                & $\pi$            \\
$\theta_2$                            & rad                               & Sin                   & 0                & $\pi$            \\
$\phi_{12}$                           & rad                               & Uniform               & 0                & $2\pi$           \\
$\phi_{\text{JL}}$                    & rad                               & Uniform               & 0                & $2\pi$           \\
$d_L$                                 & Mpc                               & Uniform Source Frame  & 20               & 1000             \\
$\theta_{\text{JN}}$                  & rad                               & Sin                   & 0                & $\pi$            \\\bottomrule
\end{tabular}
\caption{Prior distributions for the sampled parameters used in the injections.}
\label{tab:priors}
\end{table}

We run this injection on a single node with 112 cores, using the ``acceptance-walk'' variant of the dynesty implementation in the Bilby code, with five hundred live points for the nested sampling algorithm (setting {\tt nlive=500}) and a setting of  {\tt naccept=15}. These settings reduce the computational cost over typical settings for recovery of signals in noise, but were found sufficient for our purpose. We use a sampling frequency of 4096 Hz, choosing the reference frequency of the waveform to be $f_{\text{ref}}=f_{\text{ini}}=10$ Hz and starting the likelihood integration at 20 Hz, so that the modes up to $m=4$ are in band. We present in Tab.~\ref{tab:injection} the injected parameters. We select a nearly equal mass system with an edge-on orientation as this configuration maximizes the $(2,0)$ spherical harmonic mode contribution. We select a luminosity distance of 400 Mpc in order to have a high value of the signal-to-noise ratio (SNR). As the $(2,0)$ and the displacement memory contained in this and the rest of $\ell=2$ modes are subdominant effects with respect to the main $(2,\pm2)$ modes — being weaker by one to two orders of magnitude — it is necessary to account for high SNR scenarios in order to observe a significant impact in the recovery of the parameters. The network (H1 and L1) SNR for a signal with the selected parameters is 235 (the SNR for the face-on equivalent would be 470). We set the trigger time to be the one of the first GW detection, GW150914.

\begin{table}[h!]
\centering
\resizebox{\columnwidth}{!}{ 
\footnotesize
\begin{tabular}{cccccccc}
\toprule
\multicolumn{8}{c}{\textbf{Injected parameters}}   \\
\hline \hline
$\bm{m_1[M_{\odot}]}$ & $\bm{m_2[M_{\odot}]}$ & $\bm{q}$ & $\bm{\mathcal{M}[M_{\odot}]}$ & $\bm{a_{1}}$ & $\bm{a_{2}}$ & $\bm{\theta_1}$\textbf{[rad]} & $\bm{\theta_2}$\textbf{[rad]} \\ \hline
50.8                    & 49.2                    & 0.970     & 43.5                            & 0.406         & 0.271         & 0.731            & 0.729            \\ \hline \hline
\end{tabular}
}
\resizebox{\columnwidth}{!}{ 
\begin{tabular}{cccccccc}
$\bm{\phi_{\text{\textbf{12}}}}$\textbf{[rad]} & $\bm{\phi_{\text{\textbf{JL}}}}$\textbf{[rad]} & $\bm{d_L}$\textbf{[Mpc]} & $\bm{\theta_{\text{\textbf{JN}}}}$\textbf{[rad]} & $\bm{\alpha}$\textbf{[rad]} & $\bm{\delta}$\textbf{[rad]} & $\bm{\psi}$\textbf{[rad]} & $\bm{\phi}$\textbf{[rad]} \\ \hline
1.49                              & 0                                & 400                                           & $\pi/2$           &  2.77              & -0.492         & 2.71  &  5.43          \\ \bottomrule
\end{tabular}
}
\label{tab:injection}
\caption{List of injected parameters for the zero-noise injection performed in the LIGO A$^{\#}$ sensitivity \cite{Sensitivitycurves}.}
\end{table}

\begin{figure}[htp]
\includegraphics[width=0.9\columnwidth]{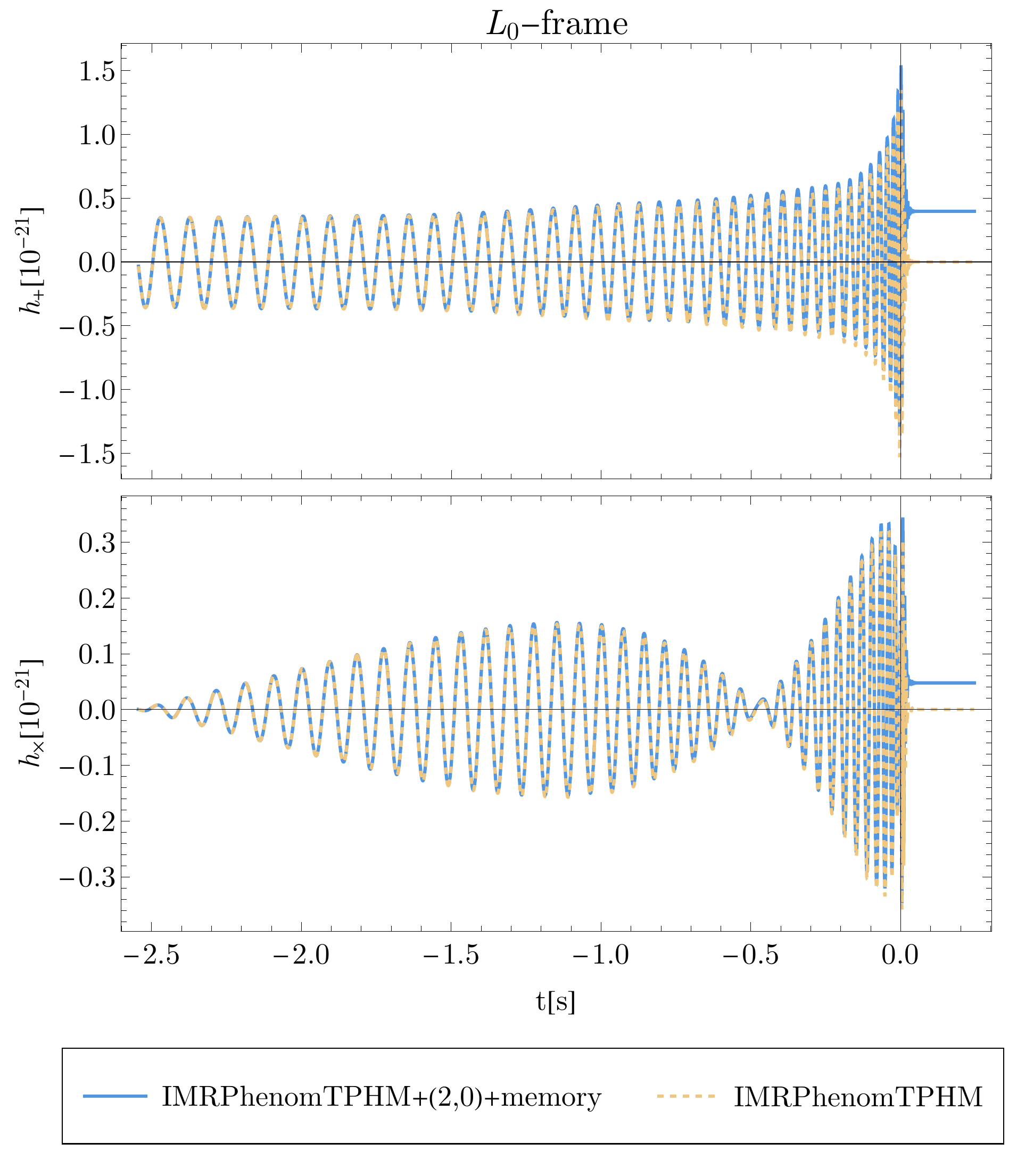}
    \caption{Waveform polarizations in the $\LO$-frame, corresponding to the parameters used for the PE injection, stated in Tab.~\ref{tab:injection}. The top panel shows the plus polarization, and the bottom panel shows the cross polarization. The blue curves correspond to the {\tt IMRPhenomTPHM} plus the addition of the $(2,0)$ mode and the memory contribution in all the $\ell=2$ modes, while the yellow dashed curves correspond to the {\tt IMRPhenomTPHM} baseline model.}
    \label{fig:WFs_inj}
\end{figure}

Fig.~\ref{fig:WFs_inj} shows the comparison of the two waveform polarizations for the parameters we used for the injection, both with and without our model of the $(2,0)$ mode added in the $\LO$-frame. For our test of PE, we inject the waveform with {\tt IMRPhenomTPHM+(2,0)+memory} and we try to recover the parameters with this same version model and only with {\tt IMRPhenomTPHM}.

In Fig.~\ref{fig:cornerplot}, we display the posterior distributions using corner plots. The one-dimensional marginal distributions of each parameter are shown along the top and right edges of the two-dimensional plots. In these plots, the black solid lines indicate the injected parameter values, while the dashed lines represent the 68\% ($1\sigma$) credible intervals of the posteriors. In the plot titles, the black value corresponds to the injected parameter, and each colored value denotes the median along with the 16th and 84th percentiles of the posterior distribution, using colors that match the distributions in the plot. In the two-dimensional distributions, the injected values are marked by a black star, and the medians of the posterior distributions are indicated with colored stars corresponding to the models used for recovery, as specified in the legends. The blue distributions correspond to the recovery with the model including the $(2,0)$ mode and the memory in all $\ell=2$ modes, while the yellow shows the recovery with the basic {\tt IMRPhenomTPHM} model without these features. In the 2D plots, the 68\%, 90\%, and 99\% credible regions are shown. The top right plot in this figure shows the relative bias between the median value of the posterior distribution and the injected value (in percentage), for each of the parameters and each version of the model.

The only quantity that has a larger bias with the full model is the individual mass of the primary object, while for the others is the simpler version model, which gives smaller biases, as expected. However, for this parameter, the posterior distributions obtained with both versions of the model look practically identical, and the small relative biases suggest that the difference is caused by statistical variations during the sampling procedure.

Overall, the posteriors with both versions of the model look very similar for all the parameters. For the mass parameters, the chirp mass is measured with high precision; however, a slight bias is observed when the $(2,0)$ mode and memory contributions are excluded, though this bias is nearly negligible. Regarding the spin parameters, a 1\% bias is present in the effective spin parameter when using only the {\tt IMRPhenomTPHM} model, while the full model accurately recovers the injected value, with a bias of only 0.09\%. The effective spin precession parameter is underestimated with both versions, and this bias increases to 9\% when the $(2,0)$ mode and memory are neglected, while when these are considered, the bias remains at 4\%. Both versions of the model provide comparable accuracy in determining and constraining the distance and inclination parameters, as the presence of higher modes in the baseline model already provides sufficient information to considerably reduce the uncertainty in these parameters. The log-likelihood increases from 27664 to 27674 when the $(2,0)$ mode and the memory are included in the recovery. This yields $\Delta\log\mathcal{L} = 10$, indicating a considerable preference for the model with the additional features, as anticipated. It is expected that the inclusion of these features in the model helps mitigate the biases further for binaries with lower total mass. However, due to computational cost, we leave additional investigations across a wider parameter space as future work.

\begin{widetext}
\begin{center}
\begin{figure}[htp]
\includegraphics[width=1\textwidth]{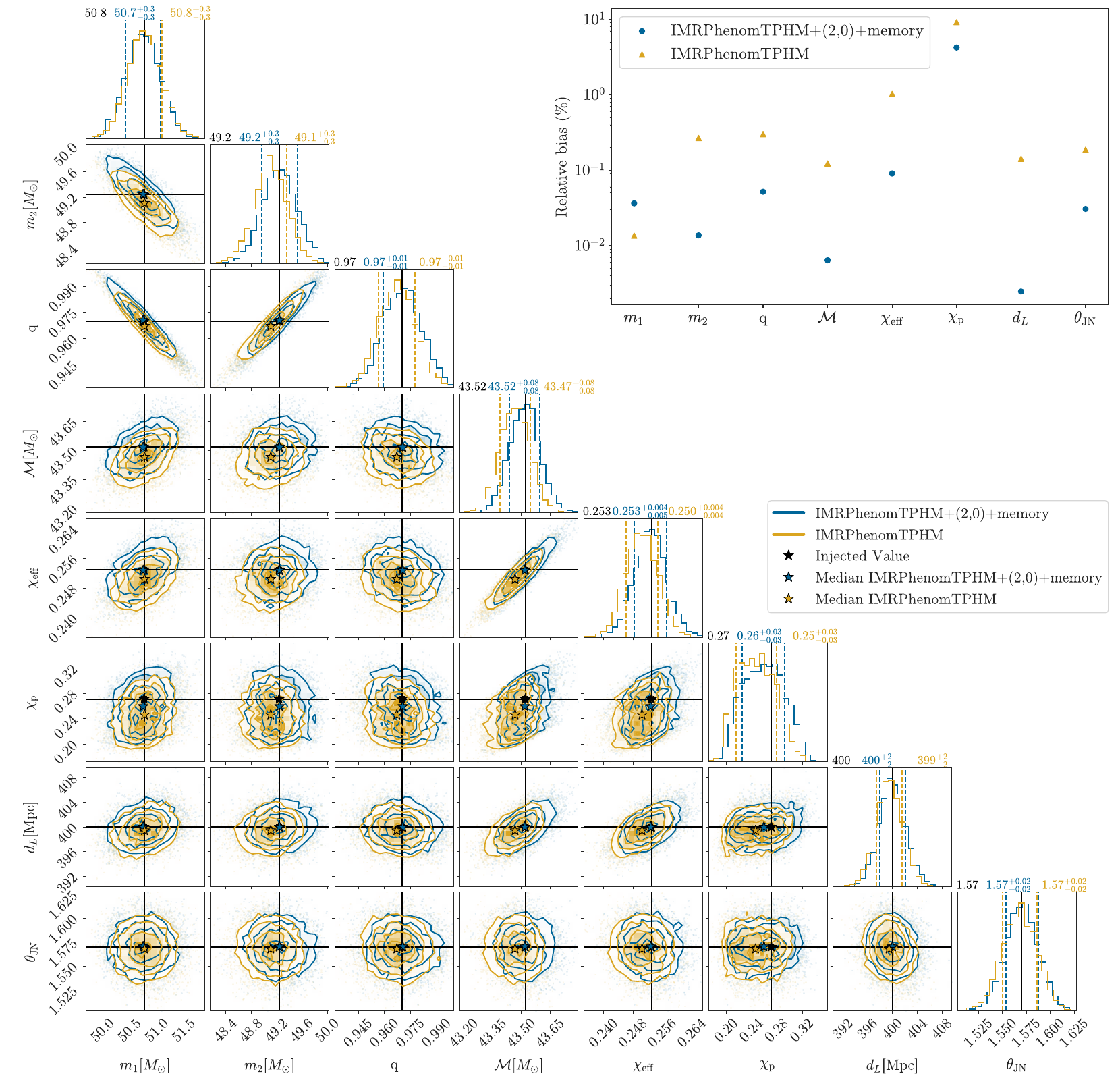}
    \caption{Full corner plot for the injection in the LIGO A$^\#$ sensitivity \cite{Sensitivitycurves}. The blue distributions represent the recovery with the model including the $(2,0)$ mode and the memory in all $\ell=2$ modes, and the yellow ones show the recovery with {\tt IMRPhenomTPHM} without these features. The parameters shown are the individual masses, the mass ratio, the chirp mass, the effective spin parameter, the effective spin precession parameter, the luminosity distance, and the inclination angle. The top right plot displays the relative bias for each of the sampled parameters and version of the model (blue dot for {\tt IMRPhenomTPHM} with the $(2,0)$ mode and memory contributions and yellow triangle for {\tt IMRPhenomTPHM}).}
    \label{fig:cornerplot}
\end{figure}
\end{center}
\end{widetext}

\section{Conclusions}
\label{sec:conclusions}
Our main result is the extension of the aligned-spin waveform model for the full $(2,0)$ spherical harmonic mode to binary systems with spin precession. We follow the same strategy as for the baseline model by treating the oscillatory and memory contributions separately. For the oscillatory component, we employ the twisting-up approximation \cite{PhysRevLett.113.151101, Schmidt:2010it, PhysRevD.86.104063} as for the rest of the modes to get them in an inertial frame using the co-precessing equivalent. For the displacement memory component, we develop analytical expressions for the memory in each of the $\ell=2$ modes using two different approaches.  In the first approach, we twist up the co-precessing modes and perform the memory integration in the inertial frame. We demonstrate that this procedure is not equivalent to first computing the memory integration and then twisting up the result, as these two operations do not commute. In the second approach, we directly take the inertial modes and compute the memory integration in the inertial frame. Finally, we check that both procedures give consistent results. We test the model accuracy by comparing the resulting waveforms with NR simulations from the SXS catalog \cite{SXS:catalog}.

This model has been implemented in the {\tt phenomxpy} Python package \cite{phenomxpy} within the {\tt IMRPhenomTPHM} waveform model \cite{Estelles:2021gvs}, in which we provide several options to enable or disable each contribution independently. Our tests confirm that including the new features developed in this work does not introduce a significant computational overhead to waveform generation.

To demonstrate the model's applicability in PE, we conduct a zero-noise injection test. We analyze the biases that arise when the full $(2,0)$ mode and memory contributions in the $\ell=2$ modes are excluded from the parameter recovery. Our results indicate that, for the specific system analyzed in LIGO A$^{\#}$ sensitivity, these biases are minor. However, they could become significant for lower-mass systems, other combinations of intrinsic parameters, or higher SNRs. Due to computational resource constraints, we leave a more extensive PE study for future work.

Future work may also include further comparisons with other waveform models using the {\tt GWMemory} package \cite{PhysRevD.98.064031} and improvements to model accuracy. Notably, since a primary source of inaccuracy arises from the Euler angles' loss of precision in the merger-ringdown regime, an NR calibration of these angles could enhance model performance.

\appendix

\section*{Acknowledgments}
The authors gratefully thank Cecilio García-Quirós for helpful comments on the manuscript. 

The authors thankfully acknowledge the computer resources at MareNostrum and the technical support provided by Barcelona Supercomputing Center (BSC) through Grant No. AECT-2024-3-0027 from the Red Española Supercomputación (RES).
M.R.-S. is supported by the Spanish Ministry of Universities via an FPU doctoral Grant (FPU21/05009).\\
This work was supported by the Universitat de les Illes Balears (UIB); the Spanish Agencia Estatal de Investigación grants PID2022-138626NB-I00, RED2022-134204-E, RED2022-134411-T, funded by MICIU/AEI/10.13039/501100011033 and by the ESF+ and the ERDF/EU; and the Comunitat Autònoma de les Illes Balears through the Conselleria d'Educació i Universitats with funds from the European Union - NextGenerationEU/PRTR-C17.I1 (SINCO2022/6719) and from the European Union - European Regional Development Fund (ERDF) (SINCO2022/18146).
\clearpage
\begin{widetext}
\section{Twisting up the modes and computation of the memory integral}
\label{appendix:twistmem}
As explained in Sec. \ref{sec:twistmem}, when applying the twisting-up approximation to the co-precessing modes, we have neglected the contribution of the time derivatives of the Euler angles that appear when we take the time derivatives of the strain. Here we provide the expression to compute the memory contribution to the $(2,0)$ spherical harmonic, when it is assumed that only the $(2,\pm2)$ co-precessing modes are included in the twisting-up approximation, and the terms involving the time derivatives of the Euler angles are not neglected.
\begin{equation}
\label{h20memeuler}
\begin{split}
    h_{2,0}^{\text{I\;mem}}&=\frac{1}{224} \sqrt{\frac{5}{6 \pi }} \int_{u_1}^u \biggl\{48 \sin ^2(\beta) \sin (4 \gamma ) \left[\left(\text{Im}\left[h_{2,2}^{\text{cp}}\right] \sin(\beta) \dot{\alpha}-\text{Re}\left[h_{2,2}^{\text{cp}}\right] \dot{\beta}\right)\left(\text{Im}\left[h_{2,2}^{\text{cp}}\right] \dot{\beta}+\text{Re}\left[h_{2,2}^{\text{cp}}\right] \sin \beta \dot{\alpha}\right)\right]\biggr.\\
    &+12 \sin ^2(\beta) \cos (4 \gamma ) \left[2\Bigl(\left(\text{Im}\left[h_{2,2}^{\text{cp}}\right]\right)^2-\left(\text{Re}\left[h_{2,2}^{\text{cp}}\right]\right)^2\Bigr)\left(\dot{\beta}^2-\sin ^2(\beta) \dot{\alpha}^2\right)+8\;\text{Im}\left[h_{2,2}^{\text{cp}}\right] \text{Re}\left[h_{2,2}^{\text{cp}}\right] \sin(\beta) \dot{\beta} \dot{\alpha}\right]\\
    &+8\;\text{Re}\left[\dot{h}_{2,2}^{\text{cp}}\right] \left[4\; \text{Im}\left[h_{2,2}^{\text{cp}}\right] (1+3 \cos (2 \beta )) \dot{\gamma}+\text{Im}\left[h_{2,2}^{\text{cp}}\right] (13 \cos(\beta) +3 \cos (3 \beta )) \dot{\alpha}-6\; \text{Re}\left[h_{2,2}^{\text{cp}}\right] \sin (2 \beta )\dot{\beta}\right]\\
    &-8\;\text{Im}\left[\dot{h}_{2,2}^{\text{cp}}\right]\left[4\; \text{Re}\left[h_{2,2}^{\text{cp}}\right] (1+3 \cos (2 \beta )) \dot{\gamma}+\text{Re}\left[h_{2,2}^{\text{cp}}\right] (13 \cos(\beta)+3 \cos (3 \beta )) \dot{\alpha}+6\; \text{Im}\left[h_{2,2}^{\text{cp}}\right] \sin (2 \beta ) \dot{\beta}\right]\\
    &+\left|h_{2,2}^{\text{cp}}\right|^2 \left[16 (13 \cos(\beta)+3 \cos (3 \beta )) \dot{\gamma}\dot{\alpha}-4 (1+3 \cos (2 \beta )) \left(\dot{\beta}^2-8 \dot{\gamma}^2\right)+(60 \cos (2 \beta )+3 \cos (4 \beta )+65) \dot{\alpha}^2\right]\\
    &\left.+8 (1+3 \cos (2 \beta )) \left|\dot{h}_{2,2}^{\text{cp}}\right|^2\right\}du.
\end{split}
\end{equation}
We show this for the $(2,0)$ spherical harmonic since this is the one that contains the most relevant contribution of the memory. It is straightforward to check that by setting $\dot{\alpha}=\dot{\beta}=\dot{\gamma}=0$ only the last term remains, which corresponds to Eq.~(\ref{twistmem20}). Now we want to show that, neglecting the contributions from these terms, we are not missing any important information from the time evolution of the waveforms. In Fig.~\ref{fig:comp_mem_terms} we plot the waveform obtained when considering the whole expression with all the terms and the one when we only include the last term $\left(8\left(3\cos(2\beta)+1\right)\left|\dot{h}_{2,2}^{\text{cp}}\right|^2\right)$ for three different NR simulations and we check that both give consistent results, therefore we can neglect the terms including the time derivatives of the Euler angles.

\begin{figure}[htp]
\includegraphics[width=1\textwidth]{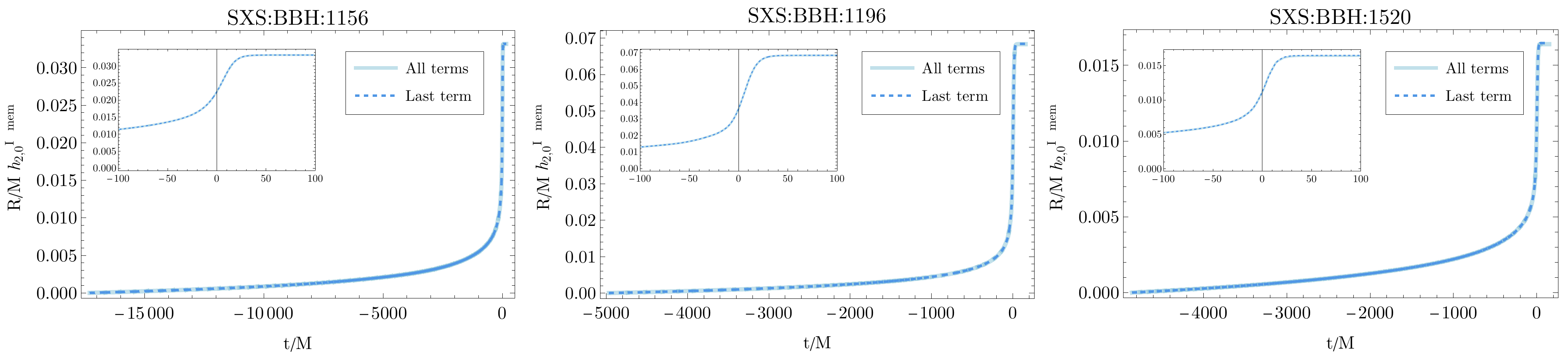}
    \caption{Comparison of the memory contribution into the $(2,0)$ mode in the $\J$-frame when including all the terms with the time derivatives of the Euler angles (light blue) and when only including the dominant term, neglecting the time derivatives of the Euler angles (dashed blue) for three different SXS simulations: SXS:BBH:1156 ($q=4.39, \bm{\chi}_1^{\text{ref}}=\{-0.142,0.228,0.381\}, \bm{\chi}_2^{\text{ref}}=\{0.314,-0.693,0.103\}, \chi_{\text{p}}=0.269$ at $Mf_{\text{ref}}=3.77\times10^{-3}$), SXS:BBH:1196 ($q=1.00, \bm{\chi}_1^{\text{ref}}=\{0.228,-0.819,0.011\}, \bm{\chi}_2^{\text{ref}}=\{0.228,-0.819,0.011\}, \chi_{\text{p}}=0.850$ at $Mf_{\text{ref}}=4.90\times10^{-3}$) and SXS:BBH:1520 ($q=3.03, \bm{\chi}_1^{\text{ref}}=\{0.540,-0.137,-0.435\}, \bm{\chi}_2^{\text{ref}}=\{0.056,0.258,0.129\}, \chi_{\text{p}}=0.557$ at $Mf_{\text{ref}}=5.15\times10^{-3}$). In the inset, we show a zoom in on the times near the merger.}
    \label{fig:comp_mem_terms}
\end{figure}

In the following equations, we present the memory contribution into each of the $\ell=2$ modes in an inertial frame computed by twisting up all the $\ell=2$ co-precessing modes, which are the ones used in the comparisons. If only the dominant $(2,\pm2)$ modes are considered in the derivation, the expressions in the set of Eqs.~(\ref{twistmem}) are recovered.

\begin{align}
\label{full_expr}
    h_{2,0}^{\text{I\;mem}} &= \frac{1}{56}\sqrt{\frac{5}{6\pi}}\int_{u_1}^u\biggl\{\left(1+3\cos(2\beta)\right)\left[2\left|\dot{h}_{2,2}^{\text{cp}}\right|^2-\left|\dot{h}_{2,1}^{\text{cp}}\right|^2-\left|\dot{h}_{2,0}^{\text{cp}}\right|^2\right]\biggr. \notag \\
    &\quad +2\cos(2\gamma)\sin^2(\beta)\left[3\left(\left(\text{Re}\left[\dot{h}_{2,1}^{\text{cp}}\right]\right)^2-\left(\text{Im}\left[\dot{h}_{2,1}^{\text{cp}}\right]\right)^2\right)+2\sqrt{6}\;\dot{h}_{2,0}^{\text{cp}}\;\text{Re}\left[\dot{h}_{2,2}^{\text{cp}}\right]\right] \notag \\
    &\quad \biggl.+2\sin(2\gamma)\sin^2(\beta)\left[6\;\text{Re}\left[\dot{h}_{2,1}^{\text{cp}}\right]\text{Im}\left[\dot{h}_{2,1}^{\text{cp}}\right]+2\sqrt{6}\;\dot{h}_{2,0}^{\text{cp}}\;\text{Im}\left[\dot{h}_{2,2}^{\text{cp}}\right]\right]\biggr\} du, \\
    h_{2,2}^{\text{I\;mem}} &= \frac{1}{112} \sqrt{\frac{5}{6 \pi }}\int_{u_1}^u e^{-2 i \alpha } \left\{-2 \sqrt{6} \sin ^2(\beta) \left[\left|\dot{h}_{2,0}^{\text{cp}}\right|^2+\left|\dot{h}_{2,1}^{\text{cp}}\right|^2-2 \left|\dot{h}_{2,2}^{\text{cp}}\right|^2\right]\right. \notag \\
    &\quad -\cos (2 \gamma ) \left[(3+\cos(2\beta)) \left(-4\;\dot{h}_{2,0}^{\text{cp}}\; \text{Re}\left[\dot{h}_{2,2}^{\text{cp}}\right]+\sqrt{6}\left[ \left(\text{Im}\left[\dot{h}_{2,1}^{\text{cp}}\right]\right)^2- \left(\text{Re}\left[\dot{h}_{2,1}^{\text{cp}}\right]\right)^2\right]\right)\right. \notag \\
    &\quad +8 i \cos(\beta) \left(2\; \dot{h}_{2,0}^{\text{cp}}\; \text{Im}\left[\dot{h}_{2,2}^{\text{cp}}\right]+\sqrt{6}\; \text{Im}\left[\dot{h}_{2,1}^{\text{cp}}\right] \text{Re}\left[\dot{h}_{2,1}^{\text{cp}}\right]\right)\biggr] \notag \\
    &\quad + 2 \sin (2 \gamma ) \biggl[(3+\cos(2\beta)) \left(2\; \dot{h}_{2,0}^{\text{cp}}\; \text{Im}\left[\dot{h}_{2,2}^{\text{cp}}\right]+\sqrt{6}\; \text{Im}\left[\dot{h}_{2,1}^{\text{cp}}\right] \text{Re}\left[\dot{h}_{2,1}^{\text{cp}}\right]\right) \notag \\
    &\quad \left.\left.+2 i \cos (\beta) \left(-4\;\dot{h}_{2,0}^{\text{cp}}\; \text{Re}\left[\dot{h}_{2,2}^{\text{cp}}\right]+\sqrt{6}\left[ \left(\text{Im}\left[\dot{h}_{2,1}^{\text{cp}}\right]\right)^2- \left(\text{Re}\left[\dot{h}_{2,1}^{\text{cp}}\right]\right)^2\right]\right)\right]\right\} du, \\
    h_{2,-2}^{\text{I\;mem}} &= \frac{1}{112} \sqrt{\frac{5}{6 \pi }}\int_{u_1}^u e^{2 i \alpha } \left\{-2 \sqrt{6} \sin ^2(\beta) \left[\left|\dot{h}_{2,0}^{\text{cp}}\right|^2+\left|\dot{h}_{2,1}^{\text{cp}}\right|^2-2 \left|\dot{h}_{2,2}^{\text{cp}}\right|^2\right]\right. \notag \\
    &\quad -\cos (2 \gamma ) \left[(3+\cos(2\beta)) \left(-4\;\dot{h}_{2,0}^{\text{cp}}\; \text{Re}\left[\dot{h}_{2,2}^{\text{cp}}\right]+\sqrt{6}\left[ \left(\text{Im}\left[\dot{h}_{2,1}^{\text{cp}}\right]\right)^2- \left(\text{Re}\left[\dot{h}_{2,1}^{\text{cp}}\right]\right)^2\right]\right)\right. \notag \\
    &\quad -8 i \cos(\beta) \left(2\;\dot{h}_{2,0}^{\text{cp}}\; \text{Im}\left[\dot{h}_{2,2}^{\text{cp}}\right]+\sqrt{6}\;\text{Im}\left[\dot{h}_{2,1}^{\text{cp}}\right] \text{Re}\left[\dot{h}_{2,1}^{\text{cp}}\right]\right)\biggr] \notag \\
    &\quad + 2 \sin (2 \gamma ) \biggl[(3+\cos(2\beta)) \left(2\;\dot{h}_{2,0}^{\text{cp}}\; \text{Im}\left[\dot{h}_{2,2}^{\text{cp}}\right]+\sqrt{6}\; \text{Im}\left[\dot{h}_{2,1}^{\text{cp}}\right] \text{Re}\left[\dot{h}_{2,1}^{\text{cp}}\right]\right) \notag \\
    &\quad \left.\left.-2 i \cos (\beta) \left(-4\;\dot{h}_{2,0}^{\text{cp}}\; \text{Re}\left[\dot{h}_{2,2}^{\text{cp}}\right]+\sqrt{6}\left[ \left(\text{Im}\left[\dot{h}_{2,1}^{\text{cp}}\right]\right)^2- \left(\text{Re}\left[\dot{h}_{2,1}^{\text{cp}}\right]\right)^2\right]\right)\right]\right\} du, \\
    h_{2,1}^{\text{I\;mem}} &= \frac{1}{28} \sqrt{\frac{5}{6 \pi }}\int_{u_1}^u e^{-i \alpha } \sin (\beta) \left\{\cos (\beta) \left[\sqrt{6} \left(\left|\dot{h}_{2,0}^{\text{cp}}\right|^2+\left|\dot{h}_{2,1}^{\text{cp}}\right|^2-2 \left|\dot{h}_{2,2}^{\text{cp}}\right|^2\right)\right.\right. \notag \\
    &\quad +\cos (2 \gamma ) \left(4\;\dot{h}_{2,0}^{\text{cp}}\; \text{Re}\left[\dot{h}_{2,2}^{\text{cp}}\right]+\sqrt{6}\left[\left(\text{Re}\left[\dot{h}_{2,1}^{\text{cp}}\right]\right)^2-\left(\text{Im}\left[\dot{h}_{2,1}^{\text{cp}}\right]\right)^2\right]\right) \notag \\
    &\quad +2 \sin (2 \gamma ) \left(2\;\dot{h}_{2,0}^{\text{cp}}\; \text{Im}\left[\dot{h}_{2,2}^{\text{cp}}\right]+\sqrt{6}\; \text{Im}\left[\dot{h}_{2,1}^{\text{cp}}\right] \text{Re}\left[\dot{h}_{2,1}^{\text{cp}}\right]\right)\biggr] \notag \\
    &\quad -i \left[\sin (2 \gamma ) \left(4\;\dot{h}_{2,0}^{\text{cp}}\; \text{Re}\left[\dot{h}_{2,2}^{\text{cp}}\right]+\sqrt{6}\left[\left(\text{Re}\left[\dot{h}_{2,1}^{\text{cp}}\right]\right)^2-\left(\text{Im}\left[\dot{h}_{2,1}^{\text{cp}}\right]\right)^2\right]^2\right)\right. \notag \\
    &\quad -2 \cos (2 \gamma ) \left(2\;\dot{h}_{2,0}^{\text{cp}}\; \text{Im}\left[\dot{h}_{2,2}^{\text{cp}}\right]+\sqrt{6}\; \text{Im}\left[\dot{h}_{2,1}^{\text{cp}}\right] \text{Re}\left[\dot{h}_{2,1}^{\text{cp}}\right]\right)\biggr]\biggr\}du, \\
    h_{2,-1}^{\text{I\;mem}} &= -\frac{1}{28} \sqrt{\frac{5}{6 \pi }}\int_{u_1}^u e^{i \alpha } \sin (\beta) \left\{\cos (\beta) \left[\sqrt{6} \left(\left|\dot{h}_{2,0}^{\text{cp}}\right|^2+\left|\dot{h}_{2,1}^{\text{cp}}\right|^2-2 \left|\dot{h}_{2,2}^{\text{cp}}\right|^2\right)\right.\right. \notag \\
    &\quad +\cos (2 \gamma ) \left(4\;\dot{h}_{2,0}^{\text{cp}} \text{Re}\left[\dot{h}_{2,2}^{\text{cp}}\right]+\sqrt{6}\left[\left(\text{Re}\left[\dot{h}_{2,1}^{\text{cp}}\right]\right)^2-\left(\text{Im}\left[\dot{h}_{2,1}^{\text{cp}}\right]\right)^2\right]\right) \notag \\
    &\quad +2 \sin (2 \gamma ) \left(2\;\dot{h}_{2,0}^{\text{cp}}\; \text{Im}\left[\dot{h}_{2,2}^{\text{cp}}\right]+\sqrt{6}\; \text{Im}\left[\dot{h}_{2,1}^{\text{cp}}\right] \text{Re}\left[\dot{h}_{2,1}^{\text{cp}}\right]\right)\biggr] \notag \\
    &\quad +i \left[\sin (2 \gamma ) \left(4\;\dot{h}_{2,0}^{\text{cp}}\; \text{Re}\left[\dot{h}_{2,2}^{\text{cp}}\right]+\sqrt{6}\left[\left(\text{Re}\left[\dot{h}_{2,1}^{\text{cp}}\right]\right)^2-\left(\text{Im}\left[\dot{h}_{2,1}^{\text{cp}}\right]\right)^2\right]^2\right)\right. \notag \\
    &\quad -2 \cos (2 \gamma ) \left(2\;\dot{h}_{2,0}^{\text{cp}}\; \text{Im}\left[\dot{h}_{2,2}^{\text{cp}}\right]+\sqrt{6}\; \text{Im}\left[\dot{h}_{2,1}^{\text{cp}}\right] \text{Re}\left[\dot{h}_{2,1}^{\text{cp}}\right]\right)\biggr]\biggr\}du.
\end{align}

\section{Computation of the memory integral in the inertial frame}
\label{appendix:integralinertial}
As explained in Sec. \ref{sec:integralinertial}, an approach to obtain the memory contribution in an inertial frame is to directly compute the integration of the inertial modes. For this purpose, we consider all the $\ell=2$ spherical harmonic modes and use the expression derived from the BMS balance laws. Since in the inertial frame we cannot assume equatorial symmetry, the modes with negative $m$ also appear in the equations.

\begin{align}
\label{expr_intertial}
    h_{2,0}^{\text{I\;mem}} &= \frac{1}{28}\sqrt{\frac{5}{6\pi}}\int_{u_1}^u\left\{2\left|\dot{h}_{2,2}^{\text{I}}\right|^2+2\left|\dot{h}_{2,-2}^{\text{I}}\right|^2 -\left|\dot{h}_{2,1}^{\text{I}}\right|^2 -\left|\dot{h}_{2,-1}^{\text{I}}\right|^2 -2\left|\dot{h}_{2,0}^{\text{I}}\right|^2 \right\}du, \\
    h_{2,2}^{\text{I\;mem}} &= \frac{1}{28}\sqrt{\frac{5}{6\pi}}\int_{u_1}^u\left\{\sqrt{6}\left(\text{Im}\left[\dot{h}_{2,1}^{\text{I}}\right]-i\text{Re}\left[\dot{h}_{2,1}^{\text{I}}\right]\right)\left(\text{Im}\left[\dot{h}_{2,-1}^{\text{I}}\right]+i\text{Re}\left[\dot{h}_{2,-1}^{\text{I}}\right]\right)\right. \notag \\
    &\quad +2\;\text{Re}\left[\dot{h}_{2,0}^{\text{I}}\right]\left[\left(\text{Re}\left[\dot{h}_{2,2}^{\text{I}}\right]+\text{Re}\left[\dot{h}_{2,-2}^{\text{I}}\right]\right)+i\left(\text{Im}\left[\dot{h}_{2,2}^{\text{I}}\right]-\text{Im}\left[\dot{h}_{2,-2}^{\text{I}}\right]\right)\right] \notag \\
    &\left.\quad +2\;\text{Im}\left[\dot{h}_{2,0}^{\text{I}}\right]\left[\left(\text{Im}\left[\dot{h}_{2,2}^{\text{I}}\right]+\text{Im}\left[\dot{h}_{2,-2}^{\text{I}}\right]\right)-i\left(\text{Re}\left[\dot{h}_{2,2}^{\text{I}}\right]-\text{Re}\left[\dot{h}_{2,-2}^{\text{I}}\right]\right)\right]\right\}du, \\
    h_{2,-2}^{\text{I\;mem}} &= \frac{1}{28}\sqrt{\frac{5}{6\pi}}\int_{u_1}^u\left\{\sqrt{6}\left(\text{Im}\left[\dot{h}_{2,1}^{\text{I}}\right]+i\text{Re}\left[\dot{h}_{2,1}^{\text{I}}\right]\right)\left(\text{Im}\left[\dot{h}_{2,-1}^{\text{I}}\right]-i\text{Re}\left[\dot{h}_{2,-1}^{\text{I}}\right]\right)\right. \notag \\
    &\quad +2\;\text{Re}\left[\dot{h}_{2,0}^{\text{I}}\right]\left[\left(\text{Re}\left[\dot{h}_{2,2}^{\text{I}}\right]+\text{Re}\left[\dot{h}_{2,-2}^{\text{I}}\right]\right)-i\left(\text{Im}\left[\dot{h}_{2,2}^{\text{I}}\right]-\text{Im}\left[\dot{h}_{2,-2}^{\text{I}}\right]\right)\right] \notag \\
    &\quad \left.+2\;\text{Im}\left[\dot{h}_{2,0}^{\text{I}}\right]\left[\left(\text{Im}\left[\dot{h}_{2,2}^{\text{I}}\right]+\text{Im}\left[\dot{h}_{2,-2}^{\text{I}}\right]\right)+i\left(\text{Re}\left[\dot{h}_{2,2}^{\text{I}}\right]-\text{Re}\left[\dot{h}_{2,-2}^{\text{I}}\right]\right)\right]\right\}du, \\
    h_{2,1}^{\text{I\;mem}} &= \frac{1}{28}\sqrt{\frac{5}{6\pi}}\int_{u_1}^u\left\{\text{Re}\left[\dot{h}_{2,0}^{\text{I}}\right]\left[\left(\text{Re}\left[\dot{h}_{2,-1}^{\text{I}}\right]-\text{Re}\left[\dot{h}_{2,1}^{\text{I}}\right]\right)-i\left(\text{Im}\left[\dot{h}_{2,-1}^{\text{I}}\right]+\text{Im}\left[\dot{h}_{2,1}^{\text{I}}\right]\right)\right]\right. \notag \\
    &\quad +\text{Im}\left[\dot{h}_{2,0}^{\text{I}}\right]\left[\left(\text{Im}\left[\dot{h}_{2,-1}^{\text{I}}\right]-\text{Im}\left[\dot{h}_{2,1}^{\text{I}}\right]\right)+i\left(\text{Re}\left[\dot{h}_{2,-1}^{\text{I}}\right]+\text{Re}\left[\dot{h}_{2,1}^{\text{I}}\right]\right)\right] \notag \\
    &\quad -\sqrt{6}\left[\left(\text{Im}\left[\dot{h}_{2,1}^{\text{I}}\right]+i\text{Re}\left[\dot{h}_{2,1}^{\text{I}}\right]\right)\left(\text{Im}\left[\dot{h}_{2,2}^{\text{I}}\right]-i\text{Re}\left[\dot{h}_{2,2}^{\text{I}}\right]\right)\right. \notag \\
    &\quad\left. -\left.\left(\text{Im}\left[\dot{h}_{2,-1}^{\text{I}}\right]-i\text{Re}\left[\dot{h}_{2,-1}^{\text{I}}\right]\right)\left(\text{Im}\left[\dot{h}_{2,-2}^{\text{I}}\right]+i\text{Re}\left[\dot{h}_{2,-2}^{\text{I}}\right]\right)\right]\right\}du, \\
    h_{2,-1}^{\text{I\;mem}} &= \frac{1}{28}\sqrt{\frac{5}{6\pi}}\int_{u_1}^u\left\{\text{Re}\left[\dot{h}_{2,0}^{\text{I}}\right]\left[\left(\text{Re}\left[\dot{h}_{2,1}^{\text{I}}\right]-\text{Re}\left[\dot{h}_{2,-1}^{\text{I}}\right]\right)-i\left(\text{Im}\left[\dot{h}_{2,1}^{\text{I}}\right]+\text{Im}\left[\dot{h}_{2,-1}^{\text{I}}\right]\right)\right]\right. \notag \\
    &\quad +\text{Im}\left[\dot{h}_{2,0}^{\text{I}}\right]\left[\left(\text{Im}\left[\dot{h}_{2,1}^{\text{I}}\right]-\text{Im}\left[\dot{h}_{2,-1}^{\text{I}}\right]\right)+i\left(\text{Re}\left[\dot{h}_{2,1}^{\text{I}}\right]+\text{Re}\left[\dot{h}_{2,-1}^{\text{I}}\right]\right)\right] \notag \\
    &\quad +\sqrt{6}\left[\left(\text{Im}\left[\dot{h}_{2,1}^{\text{I}}\right]-i\text{Re}\left[\dot{h}_{2,1}^{\text{I}}\right]\right)\left(\text{Im}\left[\dot{h}_{2,2}^{\text{I}}\right]+i\text{Re}\left[\dot{h}_{2,2}^{\text{I}}\right]\right)\right. \notag \\
    &\quad -\left.\left.\left(\text{Im}\left[\dot{h}_{2,-1}^{\text{I}}\right]+i\text{Re}\left[\dot{h}_{2,-1}^{\text{I}}\right]\right)\left(\text{Im}\left[\dot{h}_{2,-2}^{\text{I}}\right]-i\text{Re}\left[\dot{h}_{2,-2}^{\text{I}}\right]\right)\right]\right\}du.
\end{align}
These are the expressions implemented in {\tt IMRPhenomTPHM}. We take the starting integration time $u_1$ to be the starting time of the waveform.
\end{widetext}




\bibliography{final_bibliography}

\end{document}